\documentclass{jfm}
\usepackage{amsmath}
\usepackage{graphicx}
\usepackage{epstopdf, epsfig}
\usepackage{dcolumn}% Align table columns on decimal point
\usepackage{bm}% bold math
\usepackage{ stmaryrd } % for jumps
\usepackage{color}
\usepackage{float}
\usepackage[caption = false]{subfig}
\usepackage{stmaryrd}
\usepackage{url}
\usepackage{enumitem}
\usepackage[colorlinks=true,citecolor=blue,linkcolor=blue]{hyperref}

\makeatletter
\def\namedlabel#1#2{\begingroup
    #2%
    \def\@currentlabel{#2}%
    \phantomsection\label{#1}\endgroup
}
\makeatother

\usepackage{tikz}
\usetikzlibrary{matrix,fit}

\renewcommand{\vec}[1]{\boldsymbol{\mathrm{#1}}} % Vector definition
\newcommand{\ten}[1]{\boldsymbol{\mathrm{{#1}}}}
\newcommand{\pd}{\partial}

\newcommand{\ord}{\epsilon}
\newcommand{\ordest}[1]{\mathcal{O} \left( #1\right)}
\newcommand{\pdt}{\partial_t}

\shorttitle{Transfer of mass and momentum at rough and porous surfaces}
\shortauthor{U. L\={a}cis, Y. Sudhakar, S. Pasche and S. Bagheri}

\title{Transfer of mass and momentum at rough and porous surfaces}

\author{U\v{g}is L\={a}cis\aff{1}
  \corresp{\email{ugis@mech.kth.se}},
  Y. Sudhakar\aff{1,2},
  Simon Pasche\aff{1}
 \and Shervin Bagheri\aff{1}}

\affiliation{
\aff{1}Linn\'{e} FLOW centre, Department of Mechanics KTH, SE-100 44 Stockholm, Sweden\\
\aff{2}School of Mechanical Sciences, Indian Institute of Technology Goa, Ponda--403401, India
}

\begin{document}

\maketitle

\begin{abstract}
The surface texture of materials plays a critical role in wettability, turbulence and transport phenomena. In order to design surfaces for these applications, it is desirable to characterise non-smooth and porous materials by their ability to exchange mass and momentum with flowing fluids. While the underlying physics of the tangential (slip) velocity at a fluid-solid interface is well understood, the importance and treatment of normal (transpiration) velocity and normal stress is unclear. We show that, when slip velocity varies at an interface above the texture, a non-zero transpiration velocity arises from mass conservation. The ability of a given surface texture to accommodate for a normal velocity of this kind is quantified by a {\it transpiration length}. We further demonstrate that normal momentum transfer gives rise to a pressure jump.  For a porous material, the pressure jump can be characterised by so called {\it resistance coefficients}.  By solving five Stokes problems, the introduced measures of slip, transpiration and resistance can be determined for any anisotropic non-smooth surface consisting of regularly repeating geometric patterns. The proposed conditions are a subset of effective boundary conditions derived from formal multi-scale expansion. We validate and  demonstrate the physical significance of  the effective conditions on two canonical problems --  a lid-driven cavity and  a turbulent channel flow, both with non-smooth bottom surfaces.

%Coefficients in the TR model can provide a systematic approach for detailed characterisation and design of surfaces.
\end{abstract}

%\begin{keywords}
%Authors should not enter keywords on the manuscript, as these must be chosen by the author during the online submission process and will then be added during the typesetting process (see http://journals.cambridge.org/data/\linebreak[3]relatedlink/jfm-\linebreak[3]keywords.pdf for the full list)
%\end{keywords}

%========================================================================================
%\textcolor{red}{Overal ToDo
%\begin{itemize}
%\item mention Freefem++
%\item There is no active momentum transfer across the fluid-porous interface
%\item Mention at appropriate places that the formulation is empirical-free.
%\end{itemize}}
%========================================================================================

\section{Introduction}

\begin{figure}
\centering
\includegraphics[width=1.0\linewidth]{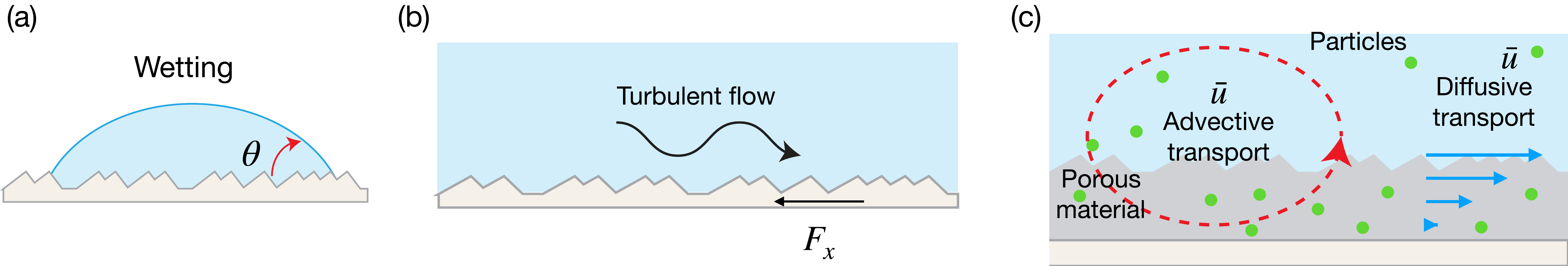}
\caption{(Colour online) Examples of problems that are sensitive to surface
properties. Droplet spreading over a rough surface (a), friction drag of turbulent
flow on a rough surface (b)
and particle transport through the interface with a rough porous material (c).}
\label{fig:intro0}
\end{figure}

The physical behaviour of a number of fluid systems is dramatically modified by the presence of a small-scale surface roughness.
For example, in wetting (figure~\ref{fig:intro0}a) -- that is when a liquid in contact with a solid reaches the balance of surface tensions -- the resulting apparent contact angle $\theta$ is very sensitive to the details of the surface texture \citep{wenzel1936resistance,quere2008wetting}.
%[cite:{Wentzel, Quere}]. 
%
%Similarly, when a solid particle sediments in a quiescent fluid very close to a wall  (figure \ref{fig:intro0},c), it will experience a drag force $F_x$ which is sensitive to the texture of the wall \citep{jeffrey1981slow,kaynan2017stokes}.
%%[cite: Jeffrey \& Oishi, Kaynan \& Yari].
%
At high Reynolds numbers (of order $1000$ and above), the pressure loss in
turbulent pipes is a function of the wall roughness (figure~\ref{fig:intro0}b) \citep{nikuradse1950laws,jimenez2004turbulent}.
%[cite: Nikuradse, Jimenez]. 
%
Yet another example is the transport phenomena involving porous media, where the exchange of mass, momentum, energy, and other passive scalars between a free flowing fluid and a porous medium depends very much on the roughness at the interface between the two domains (figure~\ref{fig:intro0}c).

%Surface wettability, surface lubrication, turbulence over rigid walls and transport phenomena in porous media, all belong to a category of fundamental problems where the large-scale observable behaviour can be dramatically modified by the presence of small-scale surface roughness. In partial wetting, i.e. when a liquid in contact with solid reaches balance of surface tensions, the resulting (apparent) Young contact angle $\theta$ (figure \ref{fig:intro0}a) is very sensitive to the details of the surface texture [cite:{Wentzel, Quere}]. When a solid sediments in a quiescent fluid very close to a wall (i.e. lubrication limit), it will experience a drag force $F_x$ (figure \ref{fig:intro0}b) which is sensitive to texture of the wall [cite: Jeffrey \& Oishi, Kaynan \& Yari]. At higher Reynolds numbers, the pressure loss in turbulent pipes is a function of wall roughness (figure \ref{fig:intro0}c) [cite: Nikuradse, Jimenez]. Finally, the exchange of of energy, mass and particles (nutritions, ions, proteins, etc) between a free flowing fluid and a porous medium depends very much on the roughness at the  interface  between the two domains (figure \ref{fig:intro0}d).

%\begin{figure}
%\centering
%\includegraphics[width=0.65\linewidth]{figs/intro0}
%\caption{(Colour online) Examples of problems that are sensitive to surface
%properties. Droplet spreading over rough surface (a), friction drag of turbulent
%flow on rough surface (b), sedimentation of particle near a rough wall (c)
%and particle transport through the interface with rough porous material (d).}
%\label{fig:intro0}
%\end{figure}

Engineers take an advantage of the sensitivity to the surface texture to modify large-scale
flow features and to enhance the transport phenomena. % in a wide range of applications. 
Efficiency of heat exchangers \citep{mehendale2000fluid,agyenim2010review}
is highly dependent on the surface texture.
%often
%depends on small surface roughness.
%leads to
%complex surface geometries, varying over small length scales.
%
In scaffold design for a bone regeneration, the cell growth on the implant (a porous biomaterial such as a calcium phosphate cement) depends on the interaction between the surrounding liquid and the surface texture of the implant \citep{dalby2007control,perez2016role}.
The performance of fuel cells depends on the ability of gas flow to efficiently transport the water vapour away from the cathode, a thin porous medium \citep{prat2002recent,haghighi2017near}.
%
%, which in turn depends on the design of the gas diffusion layer  \citep{prat2002recent,haghighi2017near}.
%
Turbulent skin friction on wings or turbine blades can be reduced by using riblets,
%careful design of complex textures such as riblets,
which are able to push quasi-streamwise vortices away from the wall \citep{walsh1984optimization}.

The design of the surface texture in the examples mentioned above is based on a
trial and error procedure, that may require tremendous amount of effort,
time and expensive surface manufacturing equipment.  
The  formulation presented in this paper provides a framework for  modelling the interaction between free flows and various textured and porous surfaces.  Our modelling approach provides a direct relationship between the microscopic geometrical details of a complex surface and the associated macroscopic transport of mass and momentum. Thus, it has the potential to replace the trial and error procedure in the design phase.

Due to the multiscale nature of the problems described above, fully resolved numerical investigations -- of both the complex surface and the free flow above it -- 
are practically impossible to perform in applied settings. 
%the microscopic simulations that requires resolving all geometric length scales are practically impossible to perform.
Effective approaches are actively pursued to circumvent this difficulty. In this way, one can 
capture the \textit{averaged} effect of the microscale features on the macroscopic processes, and hence avoid resolving microscopic geometric details.
%\revA{To ease the task, effective models have been employed for description
%of problems, if two or more distinct length scales exist in the system.}
Some recent examples of effective modelling applied to drying,
cell growth and heat exchange
can be found in works
by \cite{mosthaf2014modeling,vaca2018mechanobiological,
laloui2006experimental,wang2018transverse}.
The main challenge for effective models describing fluid-surface interaction is
the specification of a boundary condition at an artificially created interface 
% effective approach is the accurate specification of conditions at the interface
between the free-fluid region and the complex surface. Despite the recent advancements, we still lack interface conditions that capture the dominant physical features associated with complex anisotropic surfaces. 
%In this work, we provide robust and accurate interface conditions that are applicable for flows over rough as well as porous surfaces.
%This constrains efficient design of surfaces in applications.

%\revA{ \subsection{Current state-of-the-art}

\begin{figure}
\centering
\includegraphics[width=1.0\linewidth]{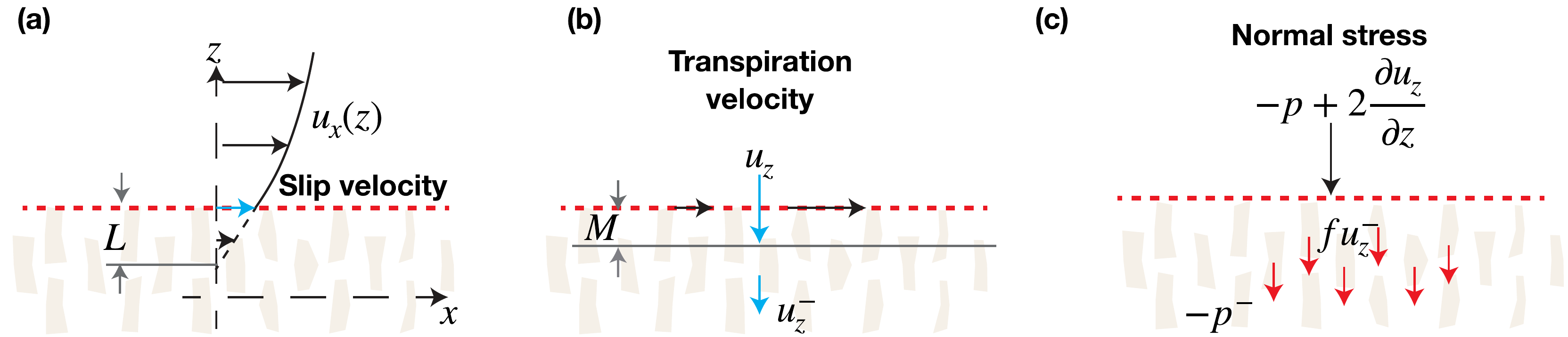}
\caption{(Colour online) Schematics of the slip length $L$ (a),
the transpiration length $M$ (b) and the resistance coefficient $f$ (c).}
\label{fig:intro1}
\end{figure}

%The interaction between the surface and the flow above mathematically and numerically is determined by a set of boundary conditions at the interface. 
%In this section, we give an overview of the existing work putting forward or examining boundary conditions  at the interface between textured or porous surface and free fluid.
Before highlighting the main ingredients of our model, we make a brief account of the current state-of-the-art of effective boundary conditions of textured and porous surfaces. 
A two-dimensional configuration is sufficient for this purpose. The streamwise and wall-normal coordinates are denoted by $x$ and $z$, 
%The main ingredients of our model can be demonstrated  on a two-dimensional simple configuration with streamwise and wall-normal coordinates $x$ and $z$, as shown in figure~\ref{fig:intro1}(a).
%In this work, we 
%assume that bulk flow within porous surface is  described by Darcy's law \citep{auriault2005filtration} and 
where the effective boundary conditions are imposed at
a planar interface at 
coordinate $z = z_i$. % the interface is aligned with the coordinate system.
For rigid textured surfaces with a characteristic size $l$, one may impose 
the slip velocity condition \citep{navier1823memoire} as an effective boundary condition,
\begin{equation}
u_x = L\, \pd_z u_x \qquad \mbox{on } z = z_i.
\label{eq:intro:slip}
\end{equation}
Here, $u_x$ is the tangential velocity component at the interface 
%above the textured surface 
and $L\sim l$ is the slip length.
%and $z$ is the interface-normal coordinate.
%
%\cite{thompson1997general} have used molecular dynamics simulations  to demonstrate the existence of slip length in nm range and its dependence of shear.  
%Furthermore, the same boundary condition applies also for textured surfaces with characteristic size $l$, which is much larger than molecular or nm scale.
%For textured surfaces with characteristic size $l$, one 
%The interaction between a textured surface and the overlaying free fluid is more
%complex.
%For such configurations, the slip length $L$ has a similar magnitude as the characteristic texture size $l$.
Geometrically -- as shown in figure~\ref{fig:intro1}(a) -- the slip length is
the distance that the velocity profile has to
be linearly extrapolated to reach zero value.
There has been extensive
development of the slip boundary condition for textured
and porous surfaces \citep{saffman1971,
sahraoui1992,miksis1994,sarkar1996,gupte1997,jager2001,
stroock2002,bolanos2017}.
In current approaches, the interface normal (transpiration) velocity
is typically set either to
zero $u_z = 0$ (for textured surfaces) or to the interior
flow $u_z = u_z^-$ (for porous surfaces)
due to mass conservation arguments or as the leading-order boundary
condition \citep{mohammadi2013pressure,lacis2016,jimenez2017derivation}.
%Physically, a large slip length
%is associated with smaller resistance imposed on a horizontal fluid
%layer just above the texture, and thus serves as a measure of exchange
%of tangential momentum. 

%,
%which leads to disappearance of symmetric part of the fluid shear
%$\pd_y u_z = 0$.

%Therefore we see that the slip length $L$ is a measure of non-smooth surfaces,
%which can be used to differentiate between various textured surfaces.
%
%Physically, it quantifies the extent to which the tangential friction
%force is modified due to the presence of texture. A
%A large slip length will also induce a large slip velocity for a given shear. 
%\revA{
%however very little work has been done on wall normal velocity.
%Different models have been proposed more specifically for turbulent flow, such
%as sand grain roughness \citep{adams2012simple}.
%
%\revA{To a certain extent, the slip length is a first valuable
%characteristic for surface design, suitable for categorisation of interaction
%between textured surface and overlaying fluid.
%However, it is rather limited, because there is no information about mass
%or normal momentum exchange.}}

%\subsubsection{Porous and poroelastic surfaces}

Configurations with porous surfaces require
additional boundary conditions. If the bulk of the surface is
governed by the Darcy-Brinkmann equation, which is typical
for works considering method of volume averaging \citep{whitaker1998method}, stress
jump conditions are often derived \citep{ochoatapia1995a,valdes2009computation,
valdes2013velocity,angot2017asymptotic}.
In the current work, however, we consider only Darcy's law 
within the bulk of the porous surface.
Consequently, a condition for the
Darcy pressure or the pore pressure $p^-$ is needed.
%To solve the flow problem, a boundary condition is needed for the
%Darcy or pore pressure $p^-$
%within the surface.
The pressure continuity $p = p^-$, where $p$ is the free
fluid pressure, has been a common choice in
the past \citep{ene1975,levy1975,hou1989,lacis2016}. The most notable recent
theoretical and numerical developments \citep{marciniak2012effective,
carraro2013pressure,carraro2018effective} have resulted in
the pressure
jump condition
\begin{equation}
%p^- - p = \mu\, C_\omega^{bl}\, \pd_z u_x \qquad \mbox{and} \qquad
p^- - p =-\mu\, C_\pi\, \pd_z u_x - 2 \mu\, \pd_z u_z.
\label{eq:intro-pjump}
\end{equation}
%for tangential flow and infiltration flow configurations, respectively. 
Here,
$\mu$ is the fluid dynamic viscosity and 
$C_\pi$ is a stabilisation parameter
derived from matching boundary layer solutions
with exterior solutions.
%can be obtained by help of help problems
%and are entirely defined by porous structure geometry and interface location
The coefficient $C_\pi$ is non-zero
only for anisotropic porous surfaces.
%
%
%The pressure condition has to be complemented with the velocity condition for the free fluid in contact with the surface. 
%The velocity boundary conditions for porous media have -- similar as the
%pressure condition -- been a subject of many investigations and are still
%
The pressure interface condition -- as well as the velocity interface condition -- for
porous media have been a subject of many investigations and is still 
debated \citep{beavers1967,han2005transmission,le2006interfacial, rosti2015direct,bottaro2016rigfibre, mikelic2000,mikelic2009,carraro2015effective, carraro2018effective,lacis2016,zampogna2019generalized}.
%different velocity conditions are used. 
%velocity condition is modified. In addition, more boundary conditions are needed
%to solve for fluid flow inside the surface.
%%

%\subsection{Scope of the present work}
%

In  this work, we extend the above conditions with new terms for the wall-normal velocity condition and the  pressure condition. Our proposed  set of boundary conditions is called {\it transpiration-resistance (TR) model,}  and it is applicable for any textured or porous surface consisting of regular repeating geometric entities. 
%The coefficients appearing in these boundary conditions provide a new way of characterising different surfaces as to their interaction with surrounding free fluid.
%The proposed set of boundary conditions is called transpiration resistance (TR) model. 
The TR model captures the transport of interface tangential momentum
 %(described by the slip boundary condition), 
 as well as the transport of mass and interface normal
momentum. It is a homogenised boundary condition, valid for configurations with a scale
separation $\ord = l/H \ll 1$, where $H$ is the characteristic length scale of the
free fluid.
%similar to ones proposed
%in literature (repeat references, or better not?).
 It
consists of the slip boundary condition \eqref{eq:intro:slip} for the interface tangential velocity. 
%
%some of the next order terms for normal velocity and
%pressure condition.
%This choice will be discussed in more details in section~\ref{sec:discuss}.
%
%
The  wall normal velocity in the TR model is 
%In this paper, we introduce 
 %
 \begin{equation}
u_z = u^-_z - M\, \pd_x u_x.
\label{eq:intro:transpiration}
\end{equation}
The first term is the seepage Darcy velocity,
%which describes influx or outflux from porous material.
%Here, $u_z$ is the wall-normal velocity 
given by $u^-_z = \left( K/\mu \right)\,\pd_z p^-$, 
%is the Darcy seepage velocity 
where $K$ is the interior permeability.
The second term quantifies how
much a surface texture allows exchange of mass with the surrounding fluid due to 
a streamwise variation of the slip velocity. %The transpiration length $M$
%and $x$ is the streamwise coordinate.
Using continuity, %  -- and 2D flow for simplicity --
the above condition for textured
surfaces (where $u^-_z=0$) can be written as $u_z = M\, \pd_z u_z$.
Geometrically (figure~\ref{fig:intro1}b), the transpiration length $M$ is thus
the distance below the interface for which a non-zero transpiration velocity
$u_z$  can exist. %-- arising from mass conservation along the interface -- can exist. 
This depth is obtained under an assumption of a linear decay of the velocity with
a slope $\pd_z u_z$.
%, in analogous way as the slip length.
%In other words, a large transpiration length is associated with a smaller resistance against exchange of mass between the free fluid and the interface region of the non-smooth surface.
%In other words, a large transpiration length is associated with smaller resistance for mass flow from the free fluid into the surface or vice versa (i.e. mass flow from within surface texture cavities to the unobstructed fluid domain). 
 %Indeed, this measure, which again is a property induced by the texture only, is an important characteristic of surfaces for controlling transport phenomena. 
%
For a porous surface, the TR model provides the pressure condition, 
%one may define additional measures arising from the
%roughness at the interface with the free flow.
%A normal force balance at the interface plane of the texture
%due to interface normal velocity $u_z$
%yields
%
 \begin{equation}
-p + 2\mu\, \pd_z u_z  =  - p^- + f\, u^-_z.
\label{eq:intro:force}
\end{equation}
Here, the left hand side is the normal stress of the outside free flow 
on the interface plane, and the right hand side is the normal
stress from the porous material.
%The right hand side is composed of the Darcian fluid pressure $p^-$ of
%the porous material and the stress borne by the solid skeleton of the porous
%material.
The resistance coefficient $f$
%%, appearing in the second term of the RHS,
quantifies
the friction force that the Darcy seepage velocity generates while passing through
the interface (figure \ref{fig:intro1}c).
%

%The proposed TR model is based on three assumptions,
%\begin{align}
%& \mbox{creeping flow assumption Re } < 1 \mbox{, which allows to solve a given flow problem} \nonumber \\
%& \mbox{near the interface with help of the linear decomposition;}  \tag{A1} \\
%& \mbox{scale separation assumption } \ord \ll 1 \mbox{, which leads to
%  constant macroscopic flow
%  field variables over the characteristic length } l \mbox{ of the surface;} \tag{A2} \\
%& \mbox{assumption that surface consists of repeating geometric entities
%  or elements, which
%  allows to consider only single structure to determine surface properties.} \tag{A3}
%\end{align}
%
%The proposed TR model is based on three assumptions,
%\begin{enumerate}
%  \item[(A1)] \label{a1} creeping flow assumption Re $< 1$, which allows to solve a given flow problem
%  near the interface with help of the linear decomposition;
%  \item[(A2)] \label{a2} scale separation assumption $\ord \ll 1$, which leads to
%  constant macroscopic flow
%  field variables over the characteristic length $l$ of the surface;
%  \item[(A3)] \label{a3} assumption that surface consists of repeating geometric entities
%  or elements, which
%  allows to consider only single structure to determine surface properties.
%\end{enumerate}
%
Three assumptions underlies the the proposed TR model.
\begin{description}[align=right,labelwidth=0.5cm]
  \item[\namedlabel{a1}{A1}]\ Creeping flow assumption Re $< 1$, which allows to solve a given flow problem
  near the interface with help of linear decomposition.
    \item[\namedlabel{a2}{A2}]\ Scale separation assumption $\ord \ll 1$, which leads to
  constant macroscopic flow
  field variables over the characteristic length $l$ of the surface.
    \item[\namedlabel{a3}{A3}]\ The surface is homogeneous i.e., it consists of repeating geometric entities
  or elements, which
  allows to consider only single structure to determine surface properties.
\end{description}

Under these assumptions, the
slip length, the transpiration length and the resistance coefficients
are properties of the surface texture only, and can be computed by solving five fundamental Stokes problems. 
% thus a valuable characteristic for surface design.
For a given texture, the knowledge of these effective coefficients provides
important information of the diffusive/advective transport into the material
as well as the ability of the solid skeleton to resist
externally imposed shear stress. The TR model is based on conditions derived from a formal multi-scale expansion in the small parameter $\epsilon$. By including higher-order terms for the transpiration velocity and for the pressure, we will show using numerical simulations that the error of the TR model is close to $\ordest{\ord^2}$.

This paper is organized as follows. In sections~\ref{sec:trm-text} and \ref{sec:poro},
we describe
and validate the TR model for textured surfaces and porous surfaces, respectively. In section~\ref{sec:turb}, we show using the turbulent channel flow that the  transpiration velocity in the TR model -- despite being a higher-order term from an asymptotic viewpoint -- is essential from a physical viewpoint.  In section~\ref{sec:discuss}, the TR model is discussed in the context of  formal multi-scale expansion and, finally, we provide conclusions in section~\ref{sec:concl}.

\section{A model for textured surfaces} \label{sec:trm-text}

In this section, we present the transpiration-resistance (TR) model for 3D textured surfaces in contact with a free flowing fluid, under the assumptions (\ref{a1}--\ref{a3}).
%(linearity, scale separation and repeating structures). 
%The  interface is flat and  aligned with the $(x,y)$ plane of the coordinate system, see figure~\ref{fig:rough-slip-decomp}.
%aligned with coordinate system in $(x,y)$ plane.
First, we
explain the boundary condition for the interface
tangential velocity (the slip condition) and show how
to obtain the associated slip length tensor. Then, we introduce the 
transpiration velocity
condition and demonstrate how to determine the transpiration length tensor
by making use of
%assumptions \ref{a1}--\ref{a3} in conjunction with 
mass conservation. Finally, we compute
the slip and transpirations tensors and validate the model by using fully resolved numerical simulations. % in a two-dimensional setting. 
The effect of the interface location 
on the accuracy of the TR model is discussed in the last subsection.

\subsection{Tangential interface velocity and slip length} \label{sec:text-slip}

The tangential velocity condition in the TR model is provided by the
standard slip condition, which for 3D textured surfaces
reads
\begin{equation}
\left( u_x, u_y \right) = \vec{u}_{t} = \frac{\ten{L}}{\mu} \cdot \vec{\tau}
 \qquad \mbox{on } z = z_i, \label{eq:text-slip}
\end{equation}
where $\vec{\tau} = \mu \left( \pd_{z} u_x, \pd_{z} u_y \right)$ 
%is vector containing dominant part of fluid shear stress in both tangential directions,
and 
$\ten{L} = \left(L_{xx}, L_{xy}; L_{xy}, L_{yy} \right)$
is the symmetric positive definite \citep{kamrin2011} surface slip length
tensor. Here, $\vec{u}_{t}$ is the tangential velocity
vector. The
tangential $t$ subscript is used interchangeably with the $x$ and $y$ components.

\begin{figure}
\centering
\includegraphics[width=0.9\linewidth]{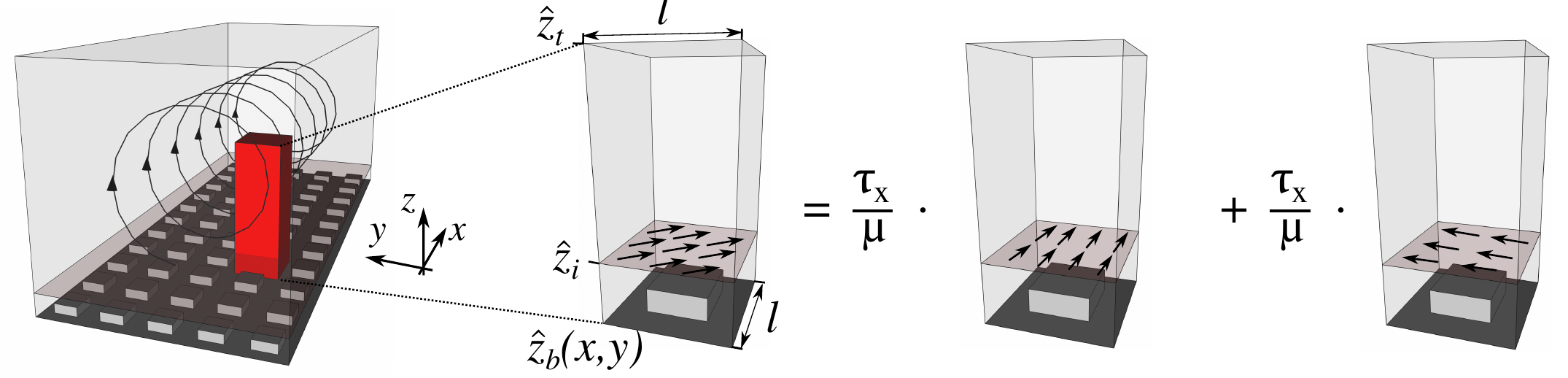}
\caption{(Colour online.) To the left, we show a flow domain with 
a generic free flow.
%one particular
%velocity distribution -- a continuous vortical structure -- is illustrated.
The flat interface with the vertical coordinate $z_i$ above the surface texture
is depicted using a transparent plane. The red rectangle is the
interface cell. To the right, we show the interface cell with a bottom
coordinate $z_b \left(x,y\right)$ -- describing the surface texture -- and
a top coordinate $z_t$.
%the linear flow field decomposition in the unit cell.
The tangential shear stress is
decomposed in unit forcing terms along the $x$ and the $y$ axis.
}
\label{fig:rough-slip-decomp}
\end{figure}

Let us consider a patterned wall and a vortical flow over it, as illustrated in
figure~\ref{fig:rough-slip-decomp}, left.
%We assume that the  length scale related to vortex ($H$)  is much larger than the roughness scale ($l$).
% The interface is located at $ z = z_{i}$.
%This scale separation 
The scale separation assumption (\ref{a2}) allows us to introduce two different spatial coordinates: $x_i$ and $\hat x_i$. The former  is used for describing spatial variations over large length scales ($x_i\sim H$).  The latter  is used to describe microscopic variations over much smaller roughness scale ($\hat x_i \sim l$).
%
%The scale separation assumption (\ref{a2}) allows us to introduce two different spatial coordinates: $x_i$ and $\hat x_i$. The former coordinate is appropriate for describing variations of  flow quantities over large (macroscopic) length scales ($x_i\sim H$).  The latter coordinate is used to describe the flow on the much smaller (microscopic) roughness scale ($\hat x_i \sim l$).
%
The effective boundary condition \eqref{eq:text-slip} is a macroscopic condition; the microscopic features of the texture are embedded in an averaged sense in the slip length tensor $\ten{L}$.

To determine $\ten{L}$, 
%
%The repeating structure (assumption A3) allows us to investigate 
%only one characteristic solid structure. We choose
we consider a small volume
near the surface of the texture with a cross section
$l \times l$. This volume contains one representative
surface structure, see figure~\ref{fig:rough-slip-decomp}.
%We refer to this volume as "interface cell". The bottom surface of this volume is  the valley of the texture, defined by the coordinate $\hat{z}_{b} \left(x,y\right)$.
%
%Here, the coordinate variable with hat is microscale coordinate. 
%
%The top surface of this volume is located in the free fluid; the top surface coordinate is set to constant $\hat{z}_t$. By testing different interface cell heights, we have determined that $\hat{z}_{t} = \hat{z}_i + 4\,l$ typically works well, which we suggest as a ``rule-of-thumb'' for practitioners.
Within this {\it interface cell}, the scale separation assumption (\ref{a2}) allows us to treat
the shear stress from the free fluid $\vec{\tau}$
as spatially constant external parameter. 
%Essentially, this means that spatial variations of $\vec{\tau}$ scale with $H$. 
Due to the creeping flow assumption (\ref{a1}), the  equations governing the
%microscale
flow response to the free fluid shear
stress  are the Stokes equations,
\begin{align}
- \nabla \hat{p} + \mu \Delta \hat{\vec{u}} & = - \delta \left( \hat{z} - \hat{z}_{i} \right) \vec{\tau}, \label{eq:itf-cell-1} \\
 \nabla \cdot \hat{\vec{u}} & = 0. \label{eq:itf-cell-2}
\end{align}
This set of equations is equivalent to a two-domain description employing
velocity continuity and stress jump at the interface (appendix~\ref{app:dirac-delta}).
%(see Appendix~\ref{app:dirac-delta} for discussion of delta function equivalence to stress jump).
%Moreover, the hat symbol on velocity and pressure denote variables and coordinates within the unit-cell problem, i.e. microscopic quantities.
%
Additionally, equations (\ref{eq:itf-cell-1}--\ref{eq:itf-cell-2}) are
the same as previously used and derived by
\cite{luchini1991,kamrin2010,luchini2013}.
%
%Equations (\ref{eq:itf-cell-1}--\ref{eq:itf-cell-2}) are augmented with
The imposed boundary conditions are no-slip
and no-penetration at the surface of the solid structure ($\hat{z} = \hat{z}_b$).
%Due to assumption \ref{a3},
We impose periodic conditions at the vertical faces of the interface cell (due to the assumption \ref{a3}). 
At the top surface of the cell, we impose zero-stress condition to keep the shear
stress at the interface as the only driving force of the problem.

The linearity assumption (\ref{a1}) allows us to
%\citep[][sections 7, 8]{leal2007advanced} 
write the solution as a product between a response operator 
$\ten{\hat{R}}_\tau$ and the free fluid shear stress,
\begin{equation}
\hat{\vec{u}} = \ten{\hat{R}}_\tau \cdot \vec{\tau} .
\label{eq:slip-decomp:0}
\end{equation}
This can be expanded as
\begin{subeqnarray}
\hat{\vec{u}} &=&
%\ten{\hat{R}}_\tau \cdot \left( \tau_x \vec{e}_x + \tau_y \vec{e}_y \right) =
\left( \ten{\hat{R}}_\tau \cdot \vec{e}_x \right)\, \tau_x + 
\left( \ten{\hat{R}}_\tau \cdot \vec{e}_y \right)\, \tau_y \\&=& 
\hat{\vec{u}}^{(\tau x)} \, \frac{\tau_x}{\mu} + 
\hat{\vec{u}}^{(\tau y)} \, \frac{\tau_y}{\mu},
\label{eq:slip-decomp}
\end{subeqnarray}
where we have defined 
%the first column of the response matrix as velocity response to the free fluid shear forcing in the $x$ direction,
\[\hat{\vec{u}}^{(\tau x)} = \mu \ten{\hat{R}}_\tau \cdot \vec{e}_x \quad\textrm{and}\quad
%correspondingly, second column as the response to the  forcing in the $y$ direction
\hat{\vec{u}}^{(\tau y)} = \mu \ten{\hat{R}}_\tau \cdot \vec{e}_y.\]
%To determine the velocity fields $\hat{\vec{u}}^{(\tau x)}$ and $\hat{\vec{u}}^{(\tau y)}$, we insert in the linear response expression for velocity (\ref{eq:slip-decomp}) and analogous expression for pressure into the governing equations for the interface cell (\ref{eq:itf-cell-1}--\ref{eq:itf-cell-2}). We demand that the resulting equations hold for any arbitrary shear stress, which gives us two fundamental problems,
Thus, the velocity fields $\hat{\vec{u}}^{(\tau x)}$ and $\hat{\vec{u}}^{(\tau y)}$ are solutions of the following two {\it fundamental problems},
\begin{align}
& - \nabla \hat{p}^{(\tau x)} + \mu \Delta \hat{\vec{u}}^{(\tau x)}
= - \mu \delta \left( \hat{z} - \hat{z}_{i} \right) \vec{e}_x, & \nabla \cdot \hat{\vec{u}}^{(\tau x)} & = 0; \tag{FP1} \label{eq:itf-fund-1} \\
& - \nabla \hat{p}^{(\tau y)} + \mu \Delta \hat{\vec{u}}^{(\tau y)}
= - \mu \delta \left( \hat{z} - \hat{z}_{i} \right) \vec{e}_y, & \nabla \cdot \hat{\vec{u}}^{(\tau y)} & = 0. \tag{FP2} \label{eq:itf-fund-2}
\end{align} 
The fundamental problems are forced in $x$ and $y$ directions, respectively, with unit shear at the plane $z_i$.  
%The boundary conditions for fundamental problems FP1 and FP2 arise due to the demand of arbitrary shear stress from conditions defined for the governing equations (\ref{eq:itf-cell-1}--\ref{eq:itf-cell-2}).
%Similar problems have been proposed by \cite{luchini1991}.

%After both fundamental problems FP1 and FP2 are solved, the slip length  tensor can be obtained by homogenising or averaging the microscale dependent flow field in the interface cell. 
Taking the surface average of
expression (\ref{eq:slip-decomp}$b$) at the interface, we obtain
\begin{equation}
\vec{u} = \langle \hat{\vec{u}} \rangle_{i} = \langle \hat{\vec{u}}^{(\tau x)}
\rangle_{i} \,  \frac{\tau_x}{\mu} + \langle \hat{\vec{u}}^{(\tau y)} \rangle_{i}
\,  \frac{\tau_y}{\mu}  \qquad \mbox{on } z = z_i. \label{eq:slip-decomp-2}
\end{equation}
No average is carried out for the free fluid shear stress, because it is constant within the interface cell (\ref{a2}).  
The surface average of an arbitrary quantity $\hat{a}$ is defined as
\begin{equation}
\langle \hat{a} \rangle_i = \langle \hat{a} \rangle_p \left( \hat{z}_i \right), \ \
\mbox{where}\ \ \langle \hat{a} \rangle_p \left( \hat{z} \right) = \frac{1}{l^2} \int\limits_{0}^{l}
\int\limits_{0}^{l} \hat{a} \left(\hat{x},\hat{y},\hat{z} \right)\,\mathit{d\hat{x}}
\,\mathit{d\hat{y}}. \label{eq:def-itf-avg}
\end{equation}
%\begin{equation}
%\langle \hat{a} \rangle \left( \hat{z} \right) = \frac{1}{l^2} \int\limits_{0}^{l}
%\int\limits_{0}^{l} \hat{a} \left(\hat{x},\hat{y},\hat{z} \right)\,\mathit{d\hat{x}}
%\,\mathit{d\hat{y}}\qquad \mbox{and} \qquad 
%a = \langle \hat{a} \rangle_{i} = \langle \hat{a} \rangle \left( \hat{z}_i \right), \label{eq:def-itf-avg}
%\end{equation}
By comparing the surface averaged
velocity in the interface cell (\ref{eq:slip-decomp-2}) with the slip
boundary condition (\ref{eq:text-slip}), we observe that the 
components of the slip length tensor
can be obtained as
\begin{equation}
L_{xx} = \langle \hat{u}^{(\tau x)}_x \rangle_{i}, \ \ 
L_{yx} = \langle \hat{u}^{(\tau x)}_y \rangle_{i}, \ \
L_{xy} = \langle \hat{u}^{(\tau y)}_x \rangle_{i}, \ \ 
L_{yy} = \langle \hat{u}^{(\tau y)}_y \rangle_{i}.
\label{eq:def-slip-comp} 
\end{equation}
In terms of the response operator, the slip tensor
becomes $\ten{L} = \mu\, \langle \ten{\hat{R}}_{\tau} \rangle_{i}$.
%Here, vector fields $\vec{\hat{u}}^{(\tau x)}$ and $\vec{\hat{u}}^{(\tau y)}$
%are solutions of fundamental problems FP1 and
%FP2 (\ref{eq:itf-fund-1}--\ref{eq:itf-fund-2}),
%respectively.
Note that dimension of the vector
fields $\vec{\hat{u}}^{(\tau x)}$ and $\vec{\hat{u}}^{(\tau y)}$ is velocity
per shear, which gives unit of length.

\begin{figure}
\centering
\includegraphics[width=0.9\linewidth]{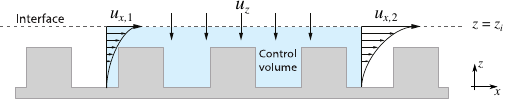}
\caption{(Colour online) Illustration of the transpiration velocity as
a consequence from mass conservation owing to the 
variation of slip velocity along the interface.
Control volume (CV) below the interface
($z = z_i$) is denoted with a shaded (blue) region.
The quantities  
$u_{x,1}$ and $u_{x,2}$ denote  slip
velocities at
the left and the right boundaries of the CV, respectively.}
\label{fig:vcv}
\end{figure}

\subsection{Interface normal velocity and transpiration length} \label{sec:text-walln}

%Before presenting the normal velocity boundary condition and the procedure to determine  the transpiration length, we provide the reader a simple motivation for the transpiration velocity based on mass conservation. 
We begin with a simple motivation for the transpiration velocity based on the mass conservation.  We consider a two dimensional rough
surface and define a control volume (CV)
below the interface $z = z_i$  as shown in figure~\ref{fig:vcv}.
We assume that there is a slip velocity
variation from $u_{x,1}$ at the left side of
the CV to $u_{x,2}$ at the right
side of the CV.
% such that $\hat{u}_{x,2} > \hat{u}_{x,1}$.
The mass fluxes % -- which is the same as volume flux for incompressible flows --
at the left and
the right boundaries of the CV are proportional to the slip velocities at the interface
(direct consequence of \ref{a1}).
Consequently, mass conservation requires a non-zero transpiration velocity $u_z$ at
the interface. If the slip velocity is increasing $u_{x,2} > u_{x,1}$, the
generated transpiration velocity is, therefore, negative.

More generally, the interface normal velocity condition in
the TR model for a 3D textured
surface is provided by a linear law relating the  normal velocity
with the tangential variation of slip velocity,
\begin{equation}
u_n = u_z = - \ten{M} : \nabla_2 \vec{u}_{t} \qquad \mbox{on } z = z_i, \label{eq:text-walln}
\end{equation}
where $\ten{M} = \left( M_{xx}, M_{xy}; M_{xy}, M_{yy} \right)$ is the
transpiration length tensor -- exhibiting the same symmetry properties as the slip
length tensor -- and $\nabla_2 = \left( \pd_x, \pd_y \right)$ is gradient operator containing 
the two tangential directions.
We use the normal $n$ subscript interchangeably with the $z$ component.
The proposed expression, motivated from the mass conservation, also emerges from a formal
multi-scale expansion (see
section~\ref{sec:discuss} and appendix~\ref{app:mse}). 
%\cite{bottaro2019adjPerspect} has also recently used the multi-scale expansion to arrive with the same transpiration velocity condition.
\cite{bottaro2019adjPerspect}  has recently used the multi-scale expansion to confirm the transpiration velocity condition proposed here.
%Such interface normal velocity can be obtained through multi-scale
%expansion (??) or motivated from mass conservation arguments.

\begin{figure}
\centering
\includegraphics[width=0.45\linewidth]{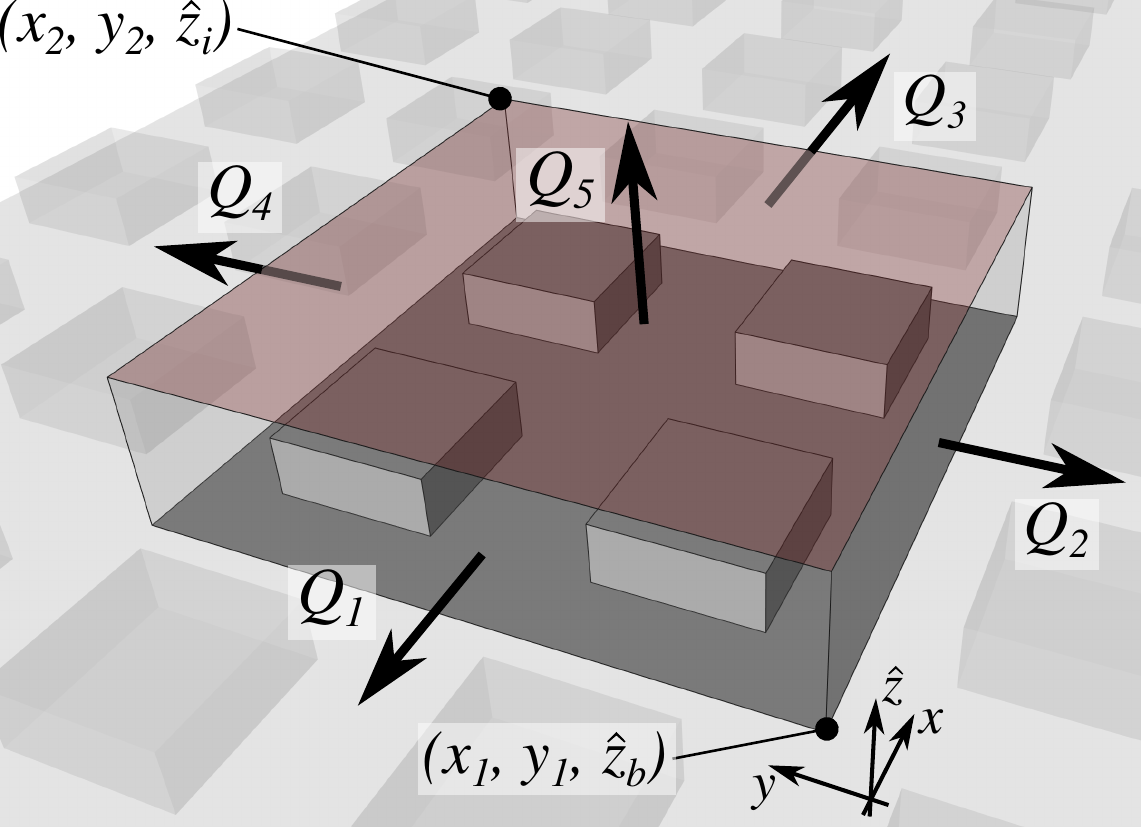}
\caption{(Color online) Control volume (CV) for deriving the transpiration
length tensor $\ten{M}$. All the possible volumetric fluxes
are indicated with thick arrows.}
\label{fig:cv-ux}
\end{figure}

To determine $\ten{M}$, we make use of mass conservation
in a 3D setting. We define a CV with size $\left( x_2 - x_1\right) \times
\left( y_2 - y_1 \right) \times \left(\hat z_i - \hat z_b \right)$ over a number of texture elements
as shown in figure~\ref{fig:cv-ux}. Taking into account that there can be no
flux through the impermeable bottom surface,  mass conservation requires
\begin{equation}
Q_1 + Q_2 + Q_3 + Q_4 + Q_5 = 0, \label{eq:all-flow-rates}
\end{equation}
where $Q_i$ are the volumetric flux through faces of the CV (figure~\ref{fig:cv-ux}).
The flux through the vertical faces ($i=1,\dots,4$) of the CV can be evaluated as
\begin{equation}
Q_i = \int_{S_i} \vec{u}\cdot\vec{n}\,\mathit{dS} = \int\limits_{\hat{z}_b}^{\hat{z}_i}
\int\limits_{s_1}^{s_2} \vec{u}\cdot\vec{n}\,\mathit{d\hat{z}}\mathit{ds},
\label{eq:Qs}
\end{equation}
where $\vec{u}$ is the effective velocity field at the CV face, $\vec{n}$ is
the unit normal vector of the surface and $s$ is either $x$ or $y$, depending on
which surface the integral is carried over. Note that the integral in the wall-normal direction is carried out over the microscale $\hat z$, because macroscopically the textured surface is infinitesimal and variations in depth does not exist.

Next, we use equation \eqref{eq:def-itf-avg} in conjunction with the solution of fundamental
problems (\ref{eq:itf-fund-1},\ref{eq:itf-fund-2}) to rewrite \eqref{eq:Qs},
% substitute the effective velocity in the integral with plane averaged interface cell solution
\begin{align}
Q_i & = \int\limits_{\hat{z}_b}^{\hat{z}_i}
\int\limits_{s_1}^{s_2} \vec{u}\cdot\vec{n}\,\mathit{d\hat{z}}\mathit{ds} = 
\int\limits_{\hat{z}_b}^{\hat{z}_i}
\int\limits_{s_1}^{s_2} \langle
\hat{\vec{u}} \rangle_p \left(x, y, \hat{z}\right)\cdot\vec{n}\,\mathit{d\hat{z}}\mathit{ds} = \\ 
& = \int\limits_{\hat{z}_b}^{\hat{z}_i}
\int\limits_{s_1}^{s_2} \left[ \langle
\hat{\ten{R}}_\tau \rangle_p \left( \hat{z} \right) \cdot \vec{\tau} \left( x,y \right)  \right]\cdot\vec{n}\,\mathit{d\hat{z}}\mathit{ds} = 
%\left( \int\limits_{\hat{z}_b}^{\hat{z}_i} \langle \ten{\hat{R}}_\tau \rangle_p\,\mathit{d\hat{z}} \cdot \int\limits_{s_1}^{s_2} \vec{\tau} \,\mathit{ds} \right) \cdot \vec{n} = 
\left( \ten{R}_\tau \cdot \int\limits_{s_1}^{s_2} \vec{\tau} \,\mathit{ds} \right) \cdot \vec{n}, \nonumber
\end{align}
%\begin{align}
%Q_i & = \int\limits_{\hat{z}_b}^{\hat{z}_i}
%\int\limits_{s_1}^{s_2} \vec{u}\cdot\vec{n}\,\mathit{d\hat{z}}\mathit{ds} = 
%\int\limits_{\hat{z}_b}^{\hat{z}_i}
%\int\limits_{s_1}^{s_2} \left( \frac{1}{l^2} \int\limits_{0}^{l} \int\limits_{0}^{l}
%\hat{\vec{u}}\,\mathit{d\hat{x}}\mathit{d\hat{y}} \right)\cdot\vec{n}\,\mathit{d\hat{z}}\mathit{ds} = \\ 
%& = \int\limits_{\hat{z}_b}^{\hat{z}_i}
%\int\limits_{s_1}^{s_2} \left( \frac{1}{l^2} \int\limits_{0}^{l} \int\limits_{0}^{l}
%\hat{\ten{R}}_\tau \left(\hat{x}, \hat{y}, \hat{z} \right) \cdot \vec{\tau} \left( x,y \right)\,\mathit{d\hat{x}}\mathit{d\hat{y}} \right)\cdot\vec{n}\,\mathit{d\hat{z}}\mathit{ds} = \left( \ten{R}_\tau \cdot \int\limits_{s_1}^{s_2} \vec{\tau} \,\mathit{ds} \right) \cdot \vec{n}. \nonumber
%\end{align}
%The final simplification emerges from commutative integral operation and by introducing a microscale volume averaged response tensor as
where we have defined the response tensor $\ten{R}_{\tau}$ as %through volume integral as
\begin{equation}
\ten{R}_{\tau} = \int\limits_{\hat{z}_b}^{\hat{z}_i} \langle \ten{\hat{R}}_\tau \rangle_p\,\mathit{d\hat{z}} = \frac{1}{l^2} \int\limits_{0}^{l} \int\limits_{0}^{l}
\int\limits_{z_{b}}^{z_{i}} \ten{\hat{R}}_{\tau} \left(\hat{x}, \hat{y}, \hat{z} \right)
\,\mathit{d\hat{x}}\,\mathit{d\hat{y}}\,\mathit{d\hat{z}}. \label{eq:def-volAvg-R}
\end{equation}
%The tensor $\ten{R}_{\tau}$ is independent of macroscale coordinates and consequently can be put outside of the integral. 
The flux through the top wall can be expressed as
\begin{equation}
Q_5 =  \int\limits_{x_1}^{x_2}  \int\limits_{y_1}^{y_2} u_z\, \mathit{dx} \mathit{dy}.
\end{equation}
Inserting the expressions for the 
fluxes through the CV faces into the mass conservation
identity (\ref{eq:all-flow-rates})
%and substituting surface unit normal vectors with corresponding coordinate unit vectors, 
we obtain
\begin{align}
 \int\limits_{x_1}^{x_2}  \int\limits_{y_1}^{y_2} u_z\, \mathit{dx} \mathit{dy} =&- \left( \ten{R}_{\tau} \cdot  \int\limits_{y_1}^{y_2} \left[
\vec{\tau} \left( x_2, y \right) - 
\vec{\tau} \left( x_1, y \right) \right] \,\mathit{dy} \right) \cdot \vec{e}_x  \nonumber \\
& - \left( \ten{R}_{\tau} \cdot  \int\limits_{x_1}^{x_2} \left[
\vec{\tau} \left( x, y_2 \right) - 
\vec{\tau} \left( x, y_1 \right) \right] \,\mathit{dx} \right) \cdot \vec{e}_y.
\end{align}
%\begin{align}
%Q_5 =& -Q_1-Q_2-Q_3-Q_4 \\
%=&- \left[ \ten{R}_{\tau} \cdot  \left( \int\limits_{y_1}^{y_2}
%\vec{\tau} \left( x_2, y \right) \,\mathit{dy} - \int\limits_{y_1}^{y_2}
%\vec{\tau} \left( x_1, y \right) \,\mathit{dy} \right) \right] \cdot \vec{e}_x - \\
%& - \left[ \ten{R}_{\tau} \cdot  \left( \int\limits_{x_1}^{x_2}
%\vec{\tau} \left( x, y_2 \right) \,\mathit{dx} - \int\limits_{x_1}^{x_2}
%\vec{\tau} \left( x, y_1 \right) \,\mathit{dx} \right) \right] \cdot \vec{e}_y.
%\end{align}
%
%Thus far we have used mass conservation and linearity to write the flux over a horizontal CV face as integral expressions of shear stress.
To proceed towards the effective boundary condition \eqref{eq:text-walln}, we take an infinitesimal CV limit, which gives
us
\begin{align}
%u_z\, \mathit{dx} \mathit{dy} =&\ - \left[ \ten{R}_{\tau} \cdot  \left(
%\vec{\tau} \left( x + \mathit{dx}, y \right) - 
%\vec{\tau} \left( x, y \right)  \right) \,\mathit{dy} \right] \cdot \vec{e}_x - \nonumber \\
%&\ - \left[ \ten{R}_{\tau} \cdot  \left(
%\vec{\tau} \left( x, y + \mathit{dy} \right) - 
%\vec{\tau} \left( x, y \right)  \right) \,\mathit{dx} \right] \cdot \vec{e}_y,
u_z\, \mathit{\Delta x} \mathit{\Delta y} =&\ - \left[ \ten{R}_{\tau} \cdot  \left(
\vec{\tau} \left( x + \mathit{ \Delta x}, y \right) - 
\vec{\tau} \left( x, y \right)  \right) \,\mathit{\Delta y} \right] \cdot \vec{e}_x  \nonumber \\
&\ - \left[ \ten{R}_{\tau} \cdot  \left(
\vec{\tau} \left( x, y + \mathit{ \Delta y} \right) - 
\vec{\tau} \left( x, y \right)  \right) \,\mathit{\Delta x} \right] \cdot \vec{e}_y,
\label{eq:Q5}
\end{align}
were we have $ \Delta y=y_2 - y_1 $, $\Delta x = x_2 - x_1 $ and  $x_1 = x$
and $y_1 = y$. Dividing both sides by $\mathit{\Delta x} \mathit{\Delta y}$ and using the definition
of a derivative, we obtain
\begin{equation}
u_z = - \left( \ten{R}_{\tau} \cdot  \pd_x \vec{\tau} \right) \cdot \vec{e}_x -
\left( \ten{R}_{\tau} \cdot  \pd_y \vec{\tau} \right)\cdot \vec{e}_y.
\end{equation} 
%
%\begin{equation}
%u_z\,\mathit{dx}\mathit{dy} = - \left[ \ten{R}_{\tau} \cdot  \left(
%\vec{\tau} \left( x_2, y \right) - 
%\vec{\tau} \left( x_1, y \right)  \right) \,\mathit{dy} \right] \cdot \vec{e}_x -
%\left[ \ten{R}_{\tau} \cdot  \left(
%\vec{\tau} \left( x, y_2 \right) - 
%\vec{\tau} \left( x, y_1 \right)  \right) \,\mathit{dx} \right] \cdot \vec{e}_y.
%\end{equation}
This expression can be rewritten using double contraction as
\begin{equation}
u_z = - \ten{R}_\tau : \nabla_2 \vec{\tau}. \label{eq:text-walln-shear}
\end{equation}
%We express
%the tangential stress with the help of the inverse slip length tensor and
%tangential velocity from (\ref{eq:text-slip}).
To obtain the transpiration
length tensor, we express the tangential shear stress
from equation (\ref{eq:text-slip}) and insert the result into
(\ref{eq:text-walln-shear}).
Comparing the final result with equation (\ref{eq:text-walln}) yields
\begin{equation}
\ten{M} = \mu \ten{R}_\tau \cdot \ten{L}^{-1}. \label{eq:def-M-tens}
\end{equation}
Recall that the tensor $\ten{R}_\tau$ can be obtained as a post-processing step from
the fundamental problems (\ref{eq:itf-fund-1}, \ref{eq:itf-fund-2}) using the volume integral (\ref{eq:def-volAvg-R}).

It is interesting to note that the velocity conditions
(\ref{eq:text-slip},\ref{eq:text-walln})
%(\ref{eq:disc-genfr1}--\ref{eq:disc-genfr2})
%for rigid rough surfaces
can be written in a more compact form, 
\begin{equation}
\left( \begin{matrix}
    u_x \\ u_y \\ u_z
\end{matrix} \right)
 = \left( \begin{matrix} 
 L_{xx}  & 0 & 0 \\
 0  & L_{xx} & 0 \\
 0 & 0 & M_{xx}
\end{matrix} \right) \cdot \left( \begin{matrix}
   \pd_z u_x \\ \pd_z u_y\\ \pd_z u_z
\end{matrix} \right), \label{eq:vel-comp-form}
\end{equation}
valid for an incompressible flow over isotropic geometries or for
 incompressible two-dimensional flows.
The upper left
$2 \times 2$ block corresponds to the slip length tensor $\ten{L}$ introduced before,
while the lower right element $M_{xx}$ is the first term of the transpiration
length tensor $\ten{M}$. 
The equivalence with the previous formulation can be seen through the application of
continuity, i.e., $\pd_z u_z = - \pd_x u_x - \pd_y u_y$. The form (\ref{eq:vel-comp-form})
can be useful in practice, for example, if boundary conditions are imposed
weakly in a finite element method. Such a set of
boundary conditions was numerically investigated by \cite{gomez2018manipulation}.
In their work, the focus was on elucidating the turbulent flow response to the
boundary condition (\ref{eq:vel-comp-form})
where all the coefficients for the slip and the
transpiration lengths could take different 
values.

\subsection{Numerical validation of velocity conditions}\label{sec:valid-lid-rough}

\begin{figure}
\centering
\hspace*{1.4cm}\subfloat[]{\includegraphics[trim = 3.5cm 21cm 11cm 2.25cm, clip, height=5.5cm]{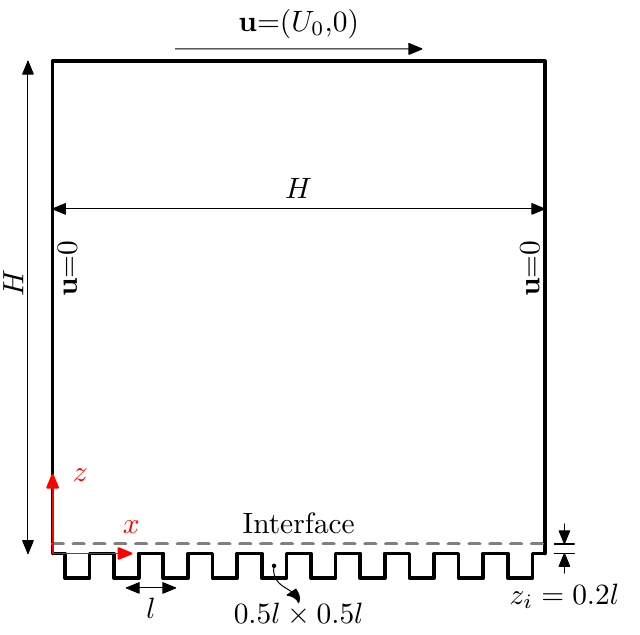}}
\hspace*{0.25cm}\subfloat[]{\includegraphics[trim = 3.5cm 21cm 11cm 2.25cm, clip, height=5.5cm]{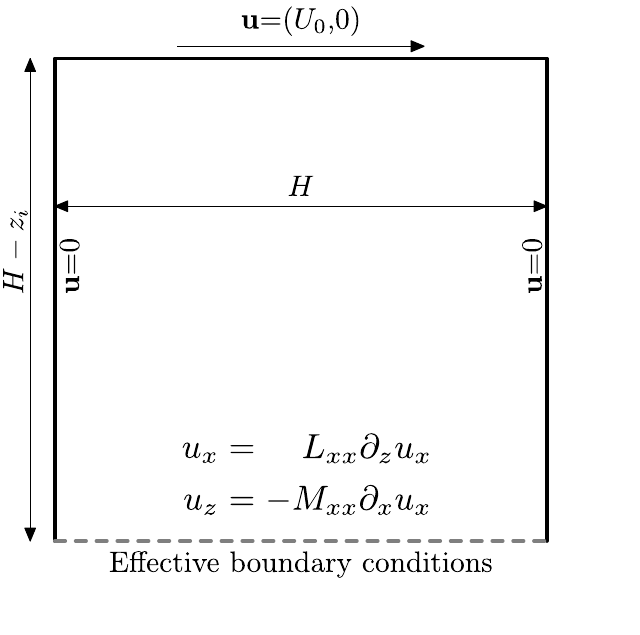}}
\caption{Lid driven cavity with a textured bottom. Frame (a) shows the computational
domain used for the resolved simulations. The bottom surface consists of
ten regular cavities. The domain for effective model simulation is
shown in (b).}
\label{fig:cavity-rough-geom}
\end{figure}

%For the validation example,
We consider a lid-driven cavity whose
bottom surface is made of a texture with the characteristic length
scale $l$ (figure~\ref{fig:cavity-rough-geom}a). The macroscopic length scale 
$H$ corresponds to the cavity length and the cavity height.
The scale separation parameter is set to $\epsilon = l/H = 0.1$.
A no-slip condition is applied
on all surfaces except the top wall, which moves with a prescribed
velocity $(U_0,0)$. 
%Direct numerical simulations (DNS) of  incompressible steady Stokes equations are carried out.
Details about the numerical solver can be found in appendix~\ref{app:lam-detail}.

%Note that while DNS is feasible here, in most practical applications, the complex nature of roughness geometries makes DNS a difficult task.

\begin{figure}
\centering
\subfloat[]{\includegraphics[trim = 0cm 0cm 0.6cm 0.65cm, clip, height=4.2cm]{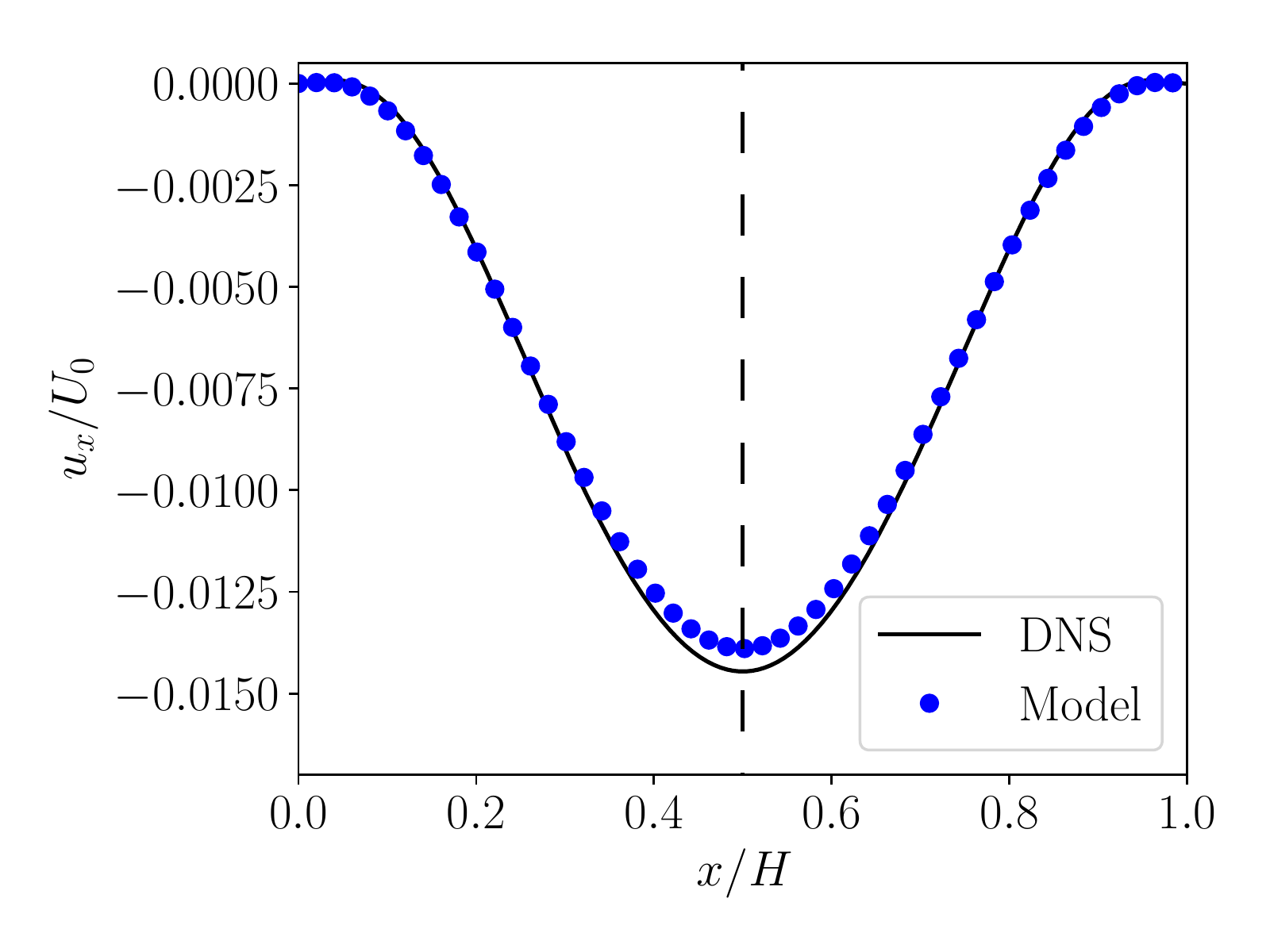}}
\subfloat[]{\includegraphics[trim = 0cm 0cm 0.6cm 0.65cm, clip, height=4.2cm]{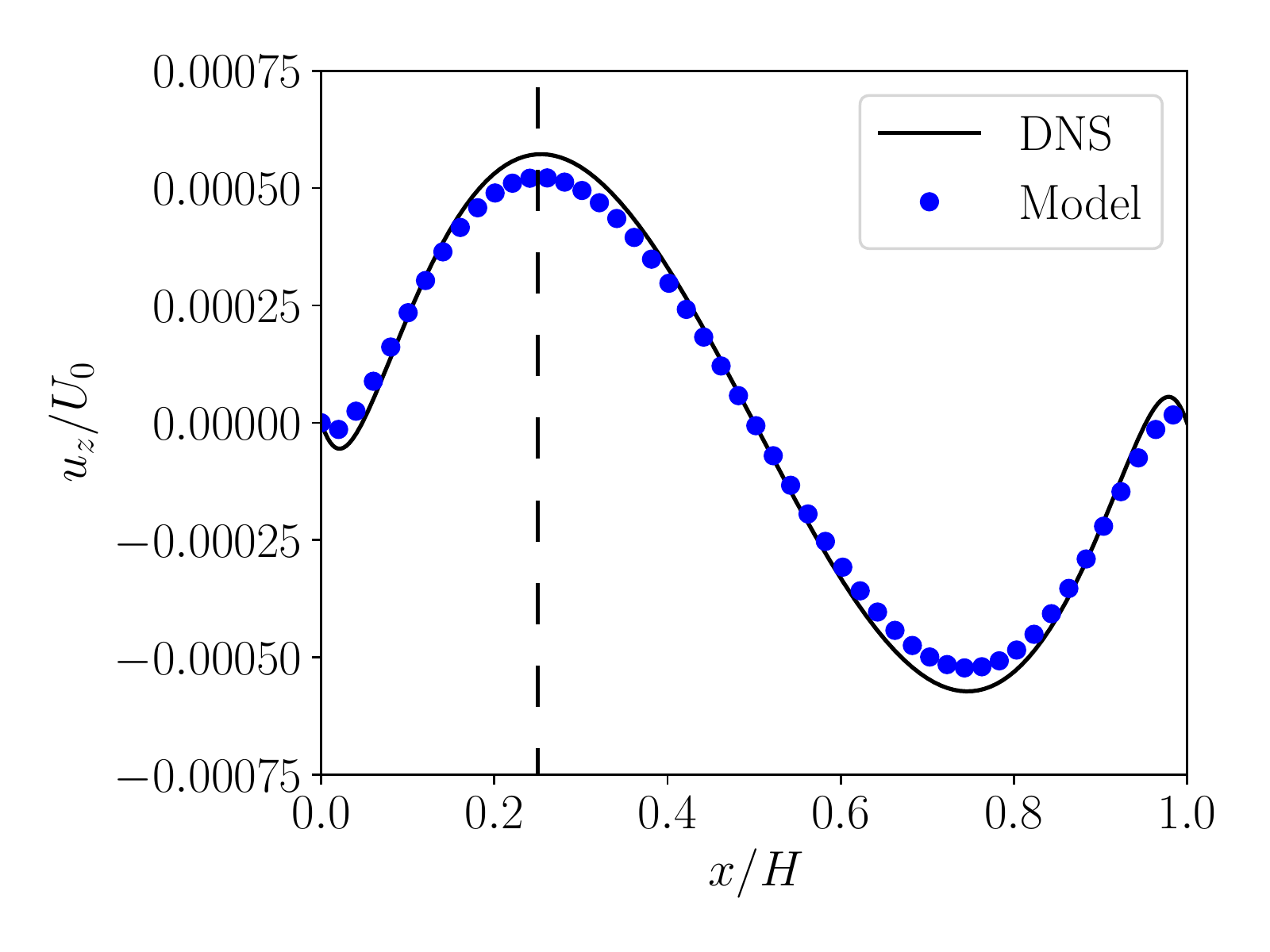}}
\caption{Tangential (a) and normal (b) velocities along the interface between the flow in the lid-driven cavity and
the rough bottom. Dashed vertical lines show the streamwise locations where the DNS and model
predictions are compared in table \ref{tab:lam-rough-itf-loc}.}
\label{fig:cavity-rough-result}
\end{figure}

The moving
upper wall generates a clock-wise rotating vortex. This vortex imposes a
negative shear on the rough surface. It also induces a downward
mass flux at the right half of the cavity and an upward
mass flux at the left half of the cavity. Near the surface texture,
one can observe velocity fluctuations with a wavelength corresponding to the
texture size $l$.
To obtain  macroscopic flow fields  from DNS, we average out the microscale oscillations by creating an ensemble of $50$ DNS simulations. The ensemble consists of configurations in
which the textured surface at the bottom of the cavity is incrementally shifted in the $x$ direction.
%In order to carry out meaningful comparison to the homogenised results, it is preferable to average these microscale oscillations. To this end, we crate an ensemble of DNS simulations, in which the textured surface at the bottom of the cavity is incrementally shifted along the $x$ direction.
%
%by  averaging over all velocity profiles  in the ensemble. 
%we  obtain an invariance with respect to the shift of the textured surface.
%
The tangential and the transpiration velocities from the ensemble averaged DNS
along $z = 0.2\,l$ are shown with black lines
in figure~\ref{fig:cavity-rough-result}(a,b),
respectively.
%In figures~\ref{fig:cavity-rough-result}(a) and (b) we show with solid black lines, respectively,
%the tangential and wall normal velocity components along location $z = 0.2\,l$
%obtained using ensemble averaged DNS. 
%We use $50$ configurations as the ensemble size.
% and arbitrarily select location  above the
%crest plane.

For comparison, we set up an effective simulation of the problem with the domain and boundary conditions shown in figure~\ref{fig:cavity-rough-geom}(b).
%on all boundaries. %and use the same FreeFEM++ solver as for the DNS.
%
%%%We have checked convergence of model simulations by using $\Delta s_t = \Delta s_b = 0.025\,l$ ($50\%$ reduction in mesh size) and observed that velocities obtained at the interfacefor interface location $z_i = 0.0$ (giving smallest velocity values) changed by less than $1\%$.
%
We position the interface at the previously selected coordinate $z_i = 0.2\,l$. 
%This  choice is arbitrary and will be discussed below. 
Coefficients for the
boundary conditions (\ref{eq:text-slip},\ref{eq:text-walln}) -- the slip and
the transpiration lengths -- 
are obtained as described in sections~\ref{sec:text-slip} and \ref{sec:text-walln},
using a FreeFEM++ open-source code \citep{github2016UgisShervin}.
% These solvers have been tested previously  \citep{lacis2016}, therefore we use the effective coefficients directly and do not discuss the solution in interface cell in details. These solvers have also been validated with respect to work of \cite{marciniak2012effective, carraro2013pressure}.
For the chosen configuration and interface location, we have 
$L_{xx}=0.218\,l$ and $M_{xx}= 0.110\,l$.
The velocities at $z=0.2 l$ from the effective simulation are compared to the ensemble averaged
DNS in figure~\ref{fig:cavity-rough-result}(a,b).
It is clear that the employed
boundary conditions accurately predict the
ensemble average (or the macroscopic variation) of both velocity components.

\begin{table}
\begin{center}
\begin{tabular}{p{12mm}p{12mm}p{12mm}p{18mm}p{18mm}p{22mm}p{22mm}}%{l|p{15mm}}
$z_i/l$ & $L_{xx}/l$ &  $M_{xx}/l$ & $u_x / U_0$ & $u_z / U_0$ & $u_x / \bar{u}_x$ & $u_z / \bar{u}_z$ \\\hline
$0.0$ & $0.018$ & $0.025$ & $-1.24 \cdot 10^{-3}$ & $1.09 \cdot 10^{-5}$ & $0.989\pm0.001$ & $0.820\pm0.087$ \\
$0.1$ & $0.118$ & $0.061$ & $-7.81 \cdot 10^{-3}$ & $1.67 \cdot 10^{-4}$ & $0.975\pm0.000$ & $0.933\pm0.001$ \\
$0.2$ & $0.218$ & $0.110$ & $-1.39 \cdot 10^{-2}$ & $5.22 \cdot 10^{-4}$ & $0.961\pm0.000$ & $0.914\pm0.001$ \\
$0.3$ & $0.318$ & $0.160$ & $-1.96 \cdot 10^{-2}$ & $1.05 \cdot 10^{-3}$ & $0.951\pm0.000$ & $0.895\pm0.001$ \\
$0.4$ & $0.418$ & $0.210$ & $-2.50 \cdot 10^{-2}$ & $1.74 \cdot 10^{-3}$ & $0.943\pm0.000$ & $0.880\pm0.001$ \\
$0.5$ & $0.518$ & $0.259$ & $-3.02 \cdot 10^{-2}$ & $2.58 \cdot 10^{-3}$ & $0.937\pm0.000$ & $0.869\pm0.001$ \\
\end{tabular}
\end{center}
\caption{The slip length $L_{xx}$ and the transpiration length $M_{xx}$
for a range of interface locations $z_i$ above the textured surface.
The effective tangential velocity
$u_x$ is sampled at $(0.5\,H,z_i)$. The effective transpiration velocity $u_z$ 
is sampled at $(0.25\,H,z_i)$.
%in horizontal
%and vertical direction, respectively,
%are provided.
The model predictions are finally normalised
using the ensemble averaged results from DNS 
$\bar{u}_x$ and $\bar{u}_z$.
% for horizontal and vertical velocity
%components, respectively.
}
\label{tab:lam-rough-itf-loc}
\end{table}%

%The location of the effective interface $z_i=0.2\,l$ considered above is arbitrary;
%\subsubsection{Interface location}
The results obtained using the TR model
are not sensitive to the exact interface location.
To show this, we repeat the previous effective computation
for a range of interface locations
$z_i = \left(0.0, 0.1, 0.3, 0.4, 0.5 \right)\,l$.
For each  interface coordinate, the  effective
coefficients $L_{xx}$ and $M_{xx}$ are
recomputed using the fundamental
problems (\ref{eq:itf-fund-1}--\ref{eq:itf-fund-2}), see table~\ref{tab:lam-rough-itf-loc}.
To quantitatively present the TR model predictions,
we select two streamwise positions at the interface plane, shown with vertical dashed
lines in figure~\ref{fig:cavity-rough-result}.
In
table~\ref{tab:lam-rough-itf-loc}, we show the model predictions of $u_x$ sampled
at point $\left(0.5\,H,z_i\right)$ for all interface locations. As the interface
moves upwards -- further away from the solid structures --, the value of the
predicted slip velocity increases due to a larger distance over which the viscous
friction can bring the velocity to the no-slip value at the wall. This effect
is correctly captured by the model through the linear increase of the slip length
$L_{xx}$ (table~\ref{tab:lam-rough-itf-loc}).
In other words, the information about the interface location is provided
to the effective model through the adjustment of the coefficients.
Similar behaviour can be observed also for the interface normal velocity
component $u_z$ and transpiration length $M_{xx}$.

%For a more quantitative comparison between effective model predictions and results  from DNS, we also sample the ensemble average DNS results at the same locations, corresponding values are denoted with $\bar{u}_x$ and $\bar{u}_z$.
In the last two columns of table~\ref{tab:lam-rough-itf-loc} we present 
the ratio between model predictions and the ensemble averaged
DNS result. 
%The ideal correspondence would correspond to $1.0$, while any deviation  corresponds to an error. While carrying investigations at different interface locations, we noticed a peculiarity in our ensemble averaged DNS results at the crest plane of the texture -- there were notable variations (with wave length of $0.002\,H$) that were not filtered out through the averaging procedure and did not reduce with mesh refinement.
The error estimates indicate uncertainty in the averaged DNS due to the presence to some unfiltered small-scale fluctuations. 
%Investigations of this effect we leave for future work, but we mark this  uncertainty in the table~\ref{tab:lam-rough-itf-loc} by adding error estimate. We consider a $\pm 0.01\,H$ interval around the sample point, we then find a maximum and a minimum ensemble averaged DNS value, which we then use to compute two different ratios between model and DNS results. The average of these two values  is considered to be the result, while the half of the difference is considered to be the error. 
As one can see, this uncertainty is significant only for transpiration velocity at coordinate $z = 0$.
From table~\ref{tab:lam-rough-itf-loc}, we observe that 
%the error for the slip velocity is overall small; 
for all interface locations the relative error ($1-u/\bar{u}$) is below $7\%$ for
the slip velocity.
The relative error for the transpiration velocity is below $14\%$. There is a trend of an increasing error as
the interface is moved upward, with an exception for interface location $z_i = 0$.
% for the wall normal velocity.
%It is likely that
This exception arises due to the large uncertainty in the reference result.
%The increasing error as the interface is moved up is a trade-off of the TR model.
Despite the trend of increasing error with interface location,
%
%, however, is the  trade-off of  the TR model; 
%The TR model is a simpler framework compared to a full formal multi-scale expansion, but the accuracy is slightly lower.
%
%While the TR model is simpler framework than a full formal multi-scale expansion,
%one has to pay with slight increase of error.
the TR model has a remarkably good accuracy taking into account that the transpiration velocity is varied over two orders of magnitude (see 5th column of table~\ref{tab:lam-rough-itf-loc}).

We have carried out similar numerical computations on equilateral triangular
surface texture and obtained the same behaviour as reported above.
This investigation shows that it is possible to adjust the interface height
over distances $\ordest{l}$ without a significant loss of accuracy.
Such invariance of interface
location has already been demonstrated numerically by \cite{lacis2016}
and theoretically by \cite{marciniak2012effective} for the slip
velocity alone. However, as a ``rule-of-thumb'', we suggest to place
the interface as close to the solid structure as possible without
intersecting the solids.  
%to reduce the volume, which is modelled by the linear response.

\section{The TR model for porous surfaces} \label{sec:poro}

In this section, we extend the TR model to 3D porous
surfaces by augmenting the set of boundary conditions from the previous section with a pressure condition. This is achieved by considering the transfer of normal momentum between the free flow region and the porous surface. 
%explain the TR model for general three dimensional porous surfaces in contact with the overlying free fluid, obeying assumptions A1--A3 (linearity, scale separation and repeating structures).
%
%The interface is flat and located in $(x,y)$ with vertical coordinate $z=z_i$.
%We present the velocity boundary conditions and argue that the  necessary coefficients are provided with the same approach as for rough surface. 
%Specifically, the model is augmented with a pressure condition and associated the resistance coefficients.
% can be derived by using assumptions A1--A3. At the end of this section, we compute the needed coefficients for boundary conditions and validate the model through comparisons between effective simulations and fully resolved simulations for three different porous geometries.

\begin{figure}
\centering
\includegraphics[width=1.0\linewidth]{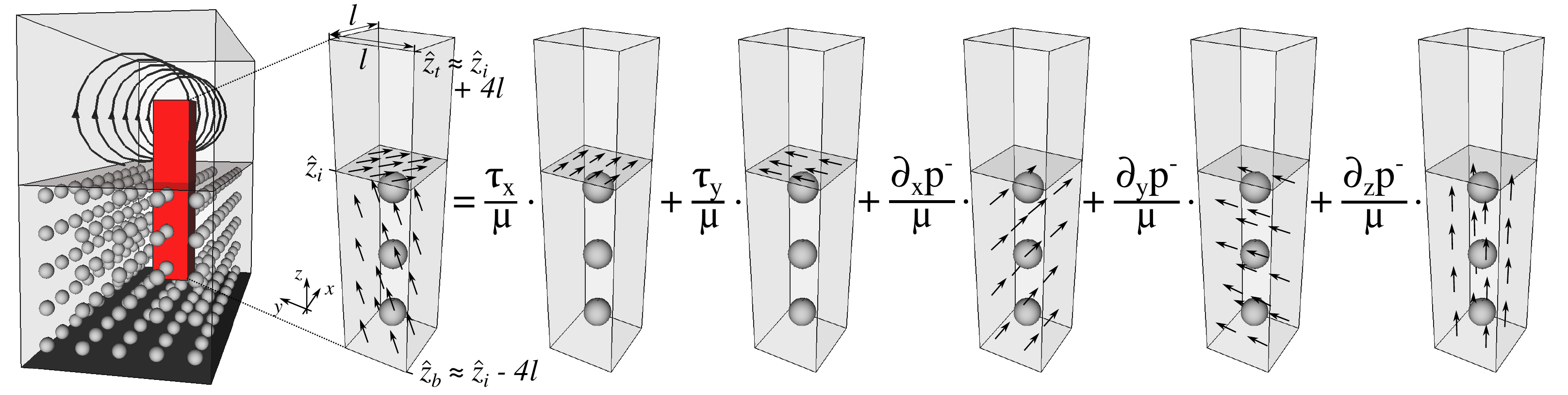}
\caption{(Colour online.) In the left frame, we illustrate a system consisting of a porous
medium and a free fluid. The transparent plane is the interface.
The solid red cuboid is the interface cell.
To the right, we show 
the interface cell and the corresponding decomposition into five fundamental problems forced either with the shear at the interface or the pore pressure gradient 
below the interface.
}
\label{fig:poro-prob-decomp}
\end{figure}

To determine the coefficients appearing in the interface boundary conditions, %the boundary conditions, 
we will adopt a similar interface-cell approach as for the textured surface
%(sections~\ref{sec:text-slip}~and~\ref{sec:text-walln}),
%
%Consider a vortical flow above a porous surface as illustrated in 
(see figure~\ref{fig:poro-prob-decomp}, leftmost frame).
%Following repeating structure assumption A3, we choose
%The interface cell near and across the porous surface with cross section $l \times l$ (figure~\ref{fig:poro-prob-decomp}$b$). 
%The upper boundary  of the interface cell  has a fixed coordinate $z_t = z_i + 4\,l$, following the same rule-of-thumb as introduced for the textured surface. 
%
The bottom coordinate of the interface cell, $\hat{z}_b$, 
%differ from the
%one introduced for the textured surface and has a constant value
%,$z_b$, 
%The value of  $z_b$ must be 
is chosen such that all flow variations
have decayed inside the porous medium, where only the interior (Darcy) flow remains.
%
%introduced by the interface between
%the porous material and free fluid have decayed and only interior (Darcy) flow remains.
%For this coordinate, we propose
As a  rule-of-thumb,  the interface cell  should contain around four solid skeleton entities $\hat{z}_b \approx \hat{z}_i - 4\,l$. From scale separation
assumption (\ref{a2}) it follows that, in the interface
cell, shear stress from the free fluid and the pore pressure gradient from the
porous material are both constant.% over the interface cell.

\subsection{Velocity boundary conditions}

For a porous surface, the tangential velocity boundary condition is identical to the textured surface, i.e. the slip condition (\ref{eq:text-slip}). 
%is defined by
%the slip length tensor
%\begin{equation}
%\vec{u}_{t} = \frac{\ten{L}}{\mu} \cdot \vec{\tau} \qquad \mbox{on } z = z_i.
%\label{eq:poro-slip}
%\end{equation}
%This boundary condition is identical to the one for the textured surfaces (\ref{eq:text-slip}). 
The interface normal velocity condition, on the other hand, is
\begin{equation}
u_n = u_n^- - \ten{M} : \nabla_2 \vec{u}_{t} \qquad 
\mbox{on } z = z_i. \label{eq:poro-walln}
\end{equation}
Here, $u_n^-$ is the interface normal Darcy velocity, 
%which in the three dimensional setting is 
%given by
satisfying
\begin{equation}
u_n^- = \left( - \frac{\ten{K}}{\mu} \cdot \nabla p^- \right) \cdot \vec{n}.
\label{eq:poro-walln-Darcy}
\end{equation}
This term is induced by
%which comes from
%The wall normal velocity is defined by transpiration length tensor $\ten{M}$ and interior permeability tensor $\ten{K}$. This boundary condition is the same as the one for the textured surfaces (\ref{eq:text-walln}), with added the interface normal Darcy velocity $u_n^-$.
%where $p^-$ is the pore pressure, as defined in the introduction. 
%The addition of the new term (compared to \eqref{eq:text-walln}) is induced by
mass conservation between free fluid and pore flow.% and provides feedback on free flow from fluid motion in porous material.

%To determine the transpiration length tensor $\ten{M}$ and the interior permeability tensor $\ten{K}$ in \eqref{eq:poro-walln} and \eqref{eq:poro-walln-Darcy}, 
%%the boundary conditions, 
%we follow a similar procedure as for the textured surface
%%(sections~\ref{sec:text-slip}~and~\ref{sec:text-walln}),
%%
%%Consider a vortical flow above a porous surface as illustrated in 
%(see figure~\ref{fig:poro-prob-decomp}$a$).
%%Following repeating structure assumption A3, we choose
%%The interface cell near and across the porous surface with cross section $l \times l$ (figure~\ref{fig:poro-prob-decomp}$b$). 
%%The upper boundary  of the interface cell  has a fixed coordinate $z_t = z_i + 4\,l$, following the same rule-of-thumb as introduced for the textured surface. 
%%
%The bottom coordinate of the interface cell
%for the porous material, $z_b$, 
%%differ from the
%%one introduced for the textured surface and has a constant value
%%,$z_b$, 
%%The value of  $z_b$ must be 
%is chosen such that all flow variations introduced by the interface between
%the porous material and free fluid have decayed and only interior (Darcy) flow remains.
%%For this coordinate, we propose
%As a  rule-of-thumb,  the interface cell  should contain around four porous material geometrical entities $z_b \approx z_i - 4\,l$. From scale separation
%assumption (A2) it then follows that in the interface
%cell, shear stress from the free fluid and the pore pressure gradient in the
%porous material are constants.% over the interface cell.

The tensors $\ten{M}$ and $\ten{L}$
are determined by solving the
 fundamental problems (\ref{eq:itf-fund-1}--\ref{eq:itf-fund-2}).
% which result in the flow response to free fluid shear forcing.
The only difference from the textured surface 
 is that the solid structures within the interface cell represent the porous
material. Consequently, all the elements in $\ten{L}$ and $\ten{M}$ can be obtained through
 expressions (\ref{eq:def-slip-comp}) and  (\ref{eq:def-M-tens}), respectively.
The interior permeability tensor~($\ten{K}$) of the porous
medium is computed through a set of Stokes
equations in a bulk unit cell \citep{whitaker1998method,meibook}.

\subsection{Pressure boundary condition} \label{sec:poro-pressure}

The pressure boundary condition for a general 3D
porous surface, obeying assumptions (\ref{a1}--\ref{a3}), is obtained through %stress balance
a balance between the normal free-fluid stress and the stress from the porous material, i.e. 
\begin{equation}
-p + 2\mu\, \pd_z u_z = -p^-
 + \vec{f}^{(1)} \cdot \vec{u}^- 
 + \vec{f}^{(2)} \cdot \vec{u}_{t} \qquad 
\mbox{on } z = z_i.
 \label{eq:poro-pres-jump}
\end{equation}
The normal stress from the porous material consists of the pore pressure $p^-$
and two friction coefficients, $\vec{f}^{(1)}$ and $\vec{f}^{(2)}$.
The coefficient $\vec{f}^{(1)}$ describes
the interface normal resistance that the Darcy flow $\vec{u}^-$ must
overcome to transport mass and momentum across
and along the interface.
%To the best of authors knowledge, this coefficient is introduced first time in literature.
The coefficient $\vec{f}^{(2)}$ provides the interface normal force due to the slip
velocity near the interface.
It exists only for anisotropic surface
geometries, similarly to the stabilisation parameter (see equation
\ref{eq:intro-pjump}) derived
by \cite{marciniak2012effective,carraro2018effective}.
%Consequently, $\vec{f}^{(2)}$ does not appear in the introductory equation (\ref{eq:intro:force}) for the TR model.

%
%The  coefficient $\vec{f}^{(2)}$ is analogous to
%the stabilisation parameters 
%%$C_\omega^{bl}$ and $C_\pi$,
%in (\ref{eq:intro-pjump}), 
%derived by .
%%To observe this analogy,
%%insert the slip velocity expression (\ref{eq:poro-slip}) and
%%the Darcy velocity expression in the
%%pressure jump and compare the result with (\ref{eq:intro-pjump}).
%The  term $\vec{f}^{(2)}$
%exists only for anisotropic surface geometries and, consequently,
%did not appear in the introductory expression (\ref{eq:intro:force}).

The friction coefficients are again determined by considering the interface
cell (figure~\ref{fig:poro-prob-decomp}).
The difference from the textured surface is the
existence of the pore pressure resulting in an additional forcing term
in the governing equations of the interface cell, yielding:
% for
%the flow in the interface cell. The governing equations are,
%due to the linearity assumption A1 are 
%the Stokes equations,
\begin{align}
- \nabla \hat{p} + \mu \Delta \hat{\vec{u}} & = - \delta \left( \hat{z} - \hat{z}_{i} \right) \vec{\tau}
+ H\left( \hat{z}_{i} - \hat{z} \right) \nabla p^-, \label{eq:itf-cell-p1} \\
 \nabla \cdot \hat{\vec{u}} & = 0. \label{eq:itf-cell-p2}
\end{align}
%where the pore pressure gradient appears as a volume forcing below
%the interface.
Darcy's law is valid only in the porous
material, therefore the pressure gradient forcing is considered only
below the interface. A 1D Heaviside step function $H\left(\hat{z}\right)$
is used to distinguish between regions above and below the 
interface. Boundary conditions for the interface cell are 
the same as for the equations (\ref{eq:itf-cell-1}--\ref{eq:itf-cell-2})
%in the interface cell of the textured surface, 
except at the bottom of the domain, where we have to impose the
interior solution corresponding
to the Darcy flow due to the same pressure gradient $\nabla p^-$
\citep{whitaker1998method,meibook,lacis2016}.

To continue, we make use of the linearity assumption (\ref{a1}) and write the
pressure as
\begin{equation}
\hat{p} = \vec{\hat{r}}_{\tau} \cdot \vec{\tau} - \vec{\hat{r}}_{p} \cdot \nabla p^-.
\end{equation}
Here,  $\vec{\hat{r}}_{\tau}$ and $\vec{\hat{r}}_{p}$ are the response operators related to the shear stress $\vec{\tau}$ and the pressure gradient $\nabla p^-$, respectively. 
%
%In the same spirit of the TR model for rough surface, 
 This expression is  expanded as
\begin{align}
\hat{p} & = \vec{\hat{r}}_{\tau} \cdot \left( \tau_x \vec{e}_x + \tau_y \vec{e}_y \right)
 - \vec{\hat{r}}_{p} \cdot \left( \pd_x p^- \vec{e}_x + \pd_y p^- \vec{e}_y + \pd_z p^- \vec{e}_z \right) = \label{eq:press-lin-resp} \\
& = \left( \vec{\hat{r}}_{\tau} \cdot \vec{e}_x \right) \tau_x + \left( \vec{\hat{r}}_{\tau} \cdot \vec{e}_y \right) \tau_y
 - \left( \vec{\hat{r}}_p \cdot \vec{e}_x \right) \pd_x p - \left( \vec{\hat{r}}_p \cdot \vec{e}_y \right) \pd_y p 
- \left( \vec{\hat{r}}_p \cdot \vec{e}_z \right) \pd_z p = \nonumber \\
& = \hat{p}^{(\tau x)} \frac{\tau_x}{\mu} + \hat{p}^{(\tau y)} \frac{\tau_y}{\mu}
- \hat{p}^{(p x)} \frac{\pd_x p}{\mu} - \hat{p}^{(p y)} \frac{\pd_y p}{\mu} 
- \hat{p}^{(p z)} \frac{\pd_z p}{\mu}, \nonumber
\end{align}
where we have defined
\begin{equation}
\hat{p}^{(\tau x)} = \mu\,\vec{\hat{r}}_{\tau} \cdot \vec{e}_x, \quad 
\hat{p}^{(\tau y)} = \mu\,\vec{\hat{r}}_{\tau} \cdot \vec{e}_y \label{eq:p-dec-tau}
\end{equation}
and
\begin{equation}
\hat{p}^{(p x)} = \mu\,\vec{\hat{r}}_{p} \cdot \vec{e}_x, \qquad
\hat{p}^{(p y)} = \mu\,\vec{\hat{r}}_{p} \cdot \vec{e}_y, \ \ \mbox{} \ \
\hat{p}^{(p z)} = \mu\,\vec{\hat{r}}_{p} \cdot \vec{e}_z. \label{eq:p-dec-pgrad}
\end{equation}
%
%After inserting this linear response (and corresponding decomposition of the velocity field) in the governing equations for the interface cell (\ref{eq:itf-cell-p1}--\ref{eq:itf-cell-p2}), we obtain five fundamental problems (FP1-FP5).
%
Note that $\hat{p}^{(\tau x)}$ and $\hat{p}^{(\tau y)}$ are the pressure
fields appearing in the fundamental problems (\ref{eq:itf-fund-1},\ref{eq:itf-fund-2}); they are the pressure responses
to the interface shear forcing in the $x$ and $y$ directions, respectively.
%
%Note that the (\ref{eq:p-dec-tau}) are pressure fields appearing in the 
%fundamental problems \ref{eq:itf-fund-1} and \ref{eq:itf-fund-2}.
%
%The first two we recover the fundamental problem  FP1 and FP2 correspond to \eqref{} and \eqref{}. of the TR model for rough surface but we make use of the pressure instead of the velocity.
%Specifically, $\hat{p}^{(\tau x)}$ and $\hat{p}^{(\tau y)}$ are the pressure responses to  interface shear forcing in the $x$ direction and
%$y$ directions, respectively.
Furthermore, $\hat{p}^{(p x)}$, $\hat{p}^{(p y)}$ and $\hat{p}^{(p z)}$ are the
pressure fields associated with the following three fundamental problems
%
%In addition, three more fundamental problems are required, which describe the response to the pressure gradient forcing
\begin{align}
 & - \nabla \hat{p}^{(p x)} + \mu \Delta \hat{\vec{u}}^{(p x)}
= - \mu H\left( \hat{z}_{i} - \hat{z} \right) \vec{e}_x, & \nabla \cdot \hat{\vec{u}}^{(p x)} & = 0; \tag{FP3} \label{eq:itf-fund-3} \\
 & - \nabla \hat{p}^{(p y)} + \mu \Delta \hat{\vec{u}}^{(p y)}
= - \mu H\left( \hat{z}_{i} - \hat{z} \right) \vec{e}_y, & \nabla \cdot \hat{\vec{u}}^{(p y)} & = 0; \tag{FP4} \label{eq:itf-fund-4} \\
 & - \nabla \hat{p}^{(p z)} + \mu \Delta \hat{\vec{u}}^{(p z)}
= - \mu H\left( \hat{z}_{i} - \hat{z} \right) \vec{e}_z, & \nabla \cdot \hat{\vec{u}}^{(p z)} & = 0. \tag{FP5} \label{eq:itf-fund-5}
\end{align}
These problems describe the response to the pressure gradient forcing along the three coordinates 
%Here, the pressure fields correspond to the response vector $\vec{\hat{r}}_p$ as
%\begin{equation}
%\hat{p}^{(p x)} = \mu\,\vec{\hat{r}}_{p} \cdot \vec{e}_x, \qquad
%\hat{p}^{(p y)} = \mu\,\vec{\hat{r}}_{p} \cdot \vec{e}_y, \ \ \mbox{and} \ \
%\hat{p}^{(p z)} = \mu\,\vec{\hat{r}}_{p} \cdot \vec{e}_z.
%\end{equation}
%The fundamental problems FP3--FP5 
and have been previously derived
by \cite{lacis2016} using formal multi-scale expansion.
Keep in mind that in equations (\ref{eq:itf-fund-3}--\ref{eq:itf-fund-5})
the fields
$\hat{\vec{u}}^{(px)}$, $\hat{\vec{u}}^{(py)}$ and $\hat{\vec{u}}^{(pz)}$ have
the dimension of length squared, similar as the permeability of a porous medium. 
%This is a different setting, if you compare with fields $\hat{\vec{u}}^{(\tau x)}$ and $\hat{\vec{u}}^{(\tau y)}$ -- these have dimensions of length only. 
%Similar distinction must be kept in mind for pressure fields in all the fundamental problems.

From the five  fundamental problems (\ref{eq:itf-fund-1}--\ref{eq:itf-fund-5}), we can determine the resistance vectors $\vec{f}^{(1)}$ and $\vec{f}^{(2)}$ for the pressure
condition (\ref{eq:poro-pres-jump}). We begin with
$\vec{f}^{(2)}$. It generates a pressure jump
\begin{equation}
p^- -p = \vec{f}^{(2)} \cdot \vec{u}_{t} = 
\vec{\tilde{f}}^{(2)} \cdot \frac{\vec{\tau}}{\mu}, \label{eq:pres-jump-sh}
\end{equation}
where $\tilde{\vec{f}}^{(2)} = \vec{f}^{(2)} \cdot \ten{L}$.
The pressure field response (\ref{eq:press-lin-resp}) due to
the shear is
\begin{equation}
\hat{p} = \hat{p}^{(\tau x)} \, \frac{\tau_x}{\mu} + \hat{p}^{(\tau y)} \,  
\frac{\tau_y}{\mu}. \label{eq:press-decomp-onlyt}
\end{equation}
Next, we need to
relate the effective pressures in porous and free fluid regions with
the linear pressure responses $\hat{p}^{(\tau x)}$ and
$\hat{p}^{(\tau y)}$ in the interface cell. For the velocity, a simple plane
average at the interface (\ref{eq:slip-decomp-2}) was sufficient. However,
for the pressure condition a single pressure value will not provide 
the necessary information
about the pressure jump. 
Therefore, we define the effective pressure
in the interior and free fluid as
\begin{equation}
p^- = \frac{1}{V_f} \int_{l^2} \int\limits_{\hat{z}_{b}}^{\hat{z}_{b}+l} \hat{p}\,\mathit{dV} 
= \langle \hat{p} \rangle^-,
\qquad p = \frac{1}{V_f} \int_{l^2} \int\limits_{\hat{z}_{t}-l}^{\hat{z}_{t}}
\hat{p}\,\mathit{dV} = \langle \hat{p} \rangle^+. \label{eq:eff-pres-def}
\end{equation}
Here, $V_f$ corresponds to fluid volume in the integration
region. % For more compact notation, we have introduced brackets.
To neglect any transition effects of the pressure field near the
interface, %-- which effective model is not intended to describe -- 
these
volume averages are taken at the bottom
and at the top of the interface cell. In this way, the averaging operation is
sufficiently far away from the interface to obtain a
representative pressure value for the interior
and the free fluid. 

Now we insert the pressure field 
decomposition (\ref{eq:press-decomp-onlyt})
into equation %in the definition of the effective  pressure 
(\ref{eq:eff-pres-def}) and we take the difference between the interior pressure and
the free fluid
pressure, 
\begin{equation}
%\vec{\tilde{f}}^{(2)} \cdot \frac{\vec{\tau}}{\mu}  = 
p^- -p =  
\left( \langle \hat{p}^{(\tau x)} \rangle^- - \langle \hat{p}^{(\tau x)} \rangle^+ \right)
\,  \frac{\tau_x}{\mu}
+ \left( \langle \hat{p}^{(\tau y)} \rangle^- 
- \langle \hat{p}^{(\tau y)} \rangle^+ \right)
\,  \frac{\tau_y}{\mu}.
\end{equation}
By comparing the above to equation (\ref{eq:pres-jump-sh}), we obtain
%From (\ref{eq:pres-jump-sh}), we get
%Requiring that this expression holds for any arbitrary shear stress value and direction, we obtain that
%
\begin{equation}
\tilde{f}^{(2)}_x = \langle \hat{p}^{(\tau x)} \rangle^- - \langle \hat{p}^{(\tau x)} \rangle^+,
\qquad
\tilde{f}^{(2)}_y = \langle \hat{p}^{(\tau y)} \rangle^- - \langle \hat{p}^{(\tau y)} \rangle^+.
\label{eq:get-F2}
\end{equation}
We emphasise that $\hat{p}^{(\tau x)}$ and $\hat{p}^{(\tau y)}$
are pressure fields in the
interface cell generated due to shear stress
forcing (figure~\ref{fig:poro-prob-decomp}, middle frame) and can be computed from the 
fundamental problems
(\ref{eq:itf-fund-1},\ref{eq:itf-fund-2}).
%These are the same fundamental problems FP1 and FP2, 
%that were used
%arising from fundamental problems previously introduced
%to determine the slip length tensor $\ten{L}$. 
Finally,
the resistance vector $\vec{f}^{(2)}$, appearing in the front of the slip velocity
in equation (\ref{eq:poro-pres-jump}), is obtained from
\begin{equation}
\vec{f}^{(2)} = \tilde{\vec{f}}^{(2)} \cdot \ten{L}^{-1}. \nonumber
\end{equation}
The procedure to get this friction coefficient is similar to the one
reported by \cite{marciniak2012effective,carraro2013pressure}.

We turn our attention to the resistance coefficient $\vec{f}^{(1)}$.
The pressure jump condition (\ref{eq:poro-pres-jump}) due to the Darcy velocity is
\begin{equation}
p^- -p = \vec{f}^{(1)} \cdot \vec{u}^- = - 
\tilde{\vec{f}}^{(1)} \cdot \frac{\nabla p^-}{\mu}, \label{eq:F1-jump-cond}
\end{equation}
where %we have introduced an auxiliaryfriction vector 
$\tilde{\vec{f}}^{(1)} = - \vec{f}^{(1)} \cdot \ten{K}$.
The pressure field response (\ref{eq:press-lin-resp}), corresponding to
the pore pressure gradient forcing, is
\begin{equation}
\hat{p} = 
- \hat{p}^{(p x)} \frac{\pd_x p^-}{\mu} - \hat{p}^{(p y)} \frac{\pd_y p^-}{\mu} 
- \hat{p}^{(p z)} \frac{\pd_z p^-}{\mu}.
\end{equation}
Using (\ref{eq:eff-pres-def}), we can express the pressure jump as
%difference between free fluid pressure and the pore pressure as
\begin{align}
& p - p^- = \langle \hat{p} \rangle^+ - \langle \hat{p} \rangle^-
= \left( \langle \hat{p}^{(px)} \rangle^- - \langle \hat{p}^{(px)} \rangle^+ \right)
\,  \frac{\pd_x p^-}{\mu} + \label{eq:F1-jump-decomp} \\ 
& + \left( \langle \hat{p}^{(py)} \rangle^- - \langle \hat{p}^{(py)} \rangle^+ \right)
\,  \frac{\pd_y p^-}{\mu}
+ \left( \langle \hat{p}^{(pz)} \rangle^- - \langle \hat{p}^{(pz)} \rangle^+ \right)
\, \frac{\pd_z p^-}{\mu}. \nonumber
\end{align}
%We compare this result with the pressure jump condition (\ref{eq:F1-jump-cond}) and
%demand that the resulting expression holds for arbitrary pore pressure gradient.
%This provides us with terms in the friction vector as
By comparing equations (\ref{eq:F1-jump-cond}) and
(\ref{eq:F1-jump-decomp}), we identify the friction vector
components as
\begin{equation}
\tilde{f}^{(1)}_x = \langle \hat{p}^{(px)} \rangle^- - \langle \hat{p}^{(px)} \rangle^+,
\ \ 
\tilde{f}^{(1)}_y = \langle \hat{p}^{(py)} \rangle^- - \langle \hat{p}^{(py)} \rangle^+,
\ \
\tilde{f}^{(1)}_z = \langle \hat{p}^{(pz)} \rangle^- - \langle \hat{p}^{(pz)} \rangle^+.
\nonumber
\end{equation}
We recall that the pressure fields in the interface cell are generated by the
pore pressure gradient forcing below
the interface (figure~\ref{fig:poro-prob-decomp}), and they are computed
from fundamental problems (\ref{eq:itf-fund-3}--\ref{eq:itf-fund-5}).
The final form of the friction coefficient is,
\begin{equation}
\vec{f}^{(1)} = - \tilde{\vec{f}}^{(1)} \cdot \ten{K}^{-1}. \nonumber
\end{equation}
This friction coefficient term, which to the best
of authors' knowledge is reported for the first time, 
is particularly important for 
capturing the correct pressure jump across the interface for layered problems,
as we demonstrate in the next section.

\subsection{Validation of the TR model for porous surfaces}\label{sec:valid-lid-poro}

\begin{figure}
\centering
\subfloat[]{\includegraphics[trim = 3.5cm 18cm 11.5cm 2.75cm, clip, height=5.5cm]{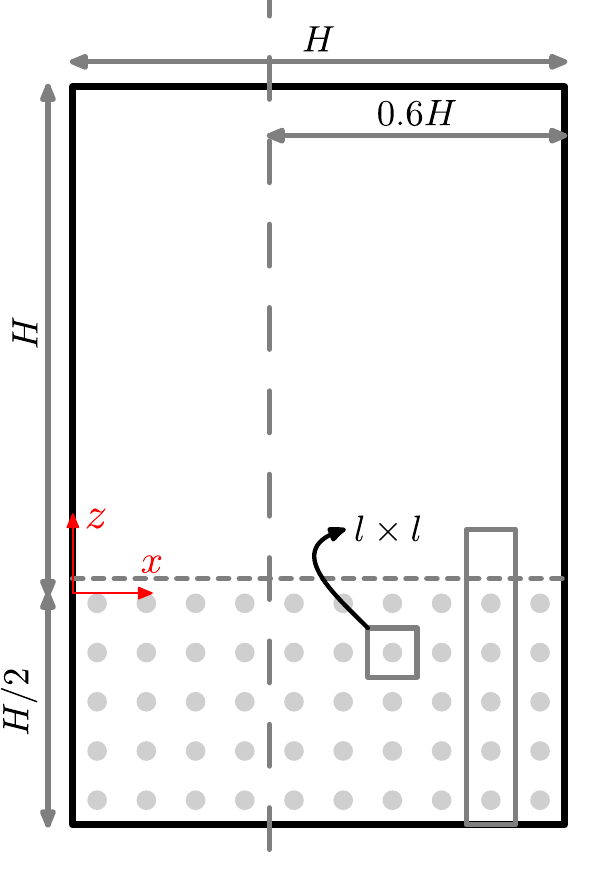}}
\hspace*{10pt}
\subfloat[]{\includegraphics[trim = 9.5cm 18cm 7cm 4cm, clip, height=5.5cm]{domain.pdf}}
\hspace*{10pt}
\subfloat[]{\includegraphics[trim = 14cm 18cm 1cm 2.75cm, clip, height=5.5cm]{domain.pdf}}
\caption{A lid-driven cavity with a porous bed. Frame (a) shows the dimensions of 
the computational domain. The dashed vertical line indicates the streamwise position where the DNS and the effective models are compared.
Frame (b) depicts an enlarged view of the porous materials showing
the microscale geometry of the three test cases considered. The
interface is located  at distance of $0.1\,l$ above the solid structure.
Frame (c) shows the domain for the continuum description.}
\label{fig:geom}
\end{figure}

We consider the same flow configuration as in section~\ref{sec:valid-lid-rough}, but we replace
the bottom textured wall with a porous surface, see figure~\ref{fig:geom}(a).
The porous medium consists of a periodic distribution of solid inclusions
with a characteristic length scale $l$ (see $l\times l$ square in 
figure~\ref{fig:geom}a).
The width and the height
of the cavity is $H$, while the depth of the porous material is $H/2$.
The scale separation parameter is again set to $\ord = l/H = 0.1$. 
%Note that the porous material contains only five repeating structures in depth; however it has been previously observed that 
The flow reaches the interior seepage velocity quickly \citep{lacis2016,lacis2016poroel};
%therefore, the porous material contains only five repeating structures in depth.
therefore, a porous material containing only five repeating structures in depth is sufficient for using Darcy equation in the interior.

To demonstrate the generality of the TR model, 
we consider three kinds of porous geometries, shown in figure~\ref{fig:geom}(b).
Configuration~(i) has circular solid
inclusions, which results in an isotropic porous medium. The anisotropic elliptic
inclusions considered in configuration~(ii) are the same
as investigated by~\cite{carraro2013pressure}.
The last geometry~(iii) has isotropic circular
inclusions with the interface layer different from the interior.
The porosity and geometrical details
of the three configurations are listed in 
table~\ref{tab:geom}. 
%Porosity $\phi$ of the medium is defined as the ratio of fluid volume to the total volume.

We carry out fully resolved DNS. 
For each configuration, the mean over an ensemble of $50$ shifted porous beds is computed. The free  flow is
similar to the flow in the lid driven cavity with the textured
bottom (section~\ref{sec:valid-lid-rough}) except that there is a mass flux in and out of the porous material.
%is able to enter and exit the porous material.
%
\begin{table}
  \begin{center}
\def~{\hphantom{0}}
  \begin{tabular}{p{23mm}ccc}%{l|p{15mm}}
    Configuration & Porosity ($\phi$) & Geometry details & Permeability tensor ($\ten{K}$) \\[10pt]
       (i) & 0.75 & $r=0.28\,l$ & $\left(\begin{array}{cc} 0.014 & 0\\ 0 & 0.014 \end{array}\right)\,l^2$ \\[0.5cm]
       (ii)   & 0.78   & $\begin{array}{c} a=0.36\,l,\ \ b=0.19\,l,  \\ \alpha=45^o \end{array}$  & $\left(\begin{array}{cc} 0.016 & 0.003\\ 0.003 & 0.016 \end{array}\right)\,l^2$\\[0.5cm]
%       (iii)  & $\begin{array}{c} \phi_i = 0.95 \\ \phi_1 = 0.66 \end{array}$  & $r_i =0.13\,l$, $r_1=0.33\,l$ & $\left(\begin{array}{cc} 0.064 & 0\\ 0 & 0.064 \end{array}\right)\,l^2$
       (iii)  & $\begin{array}{c} \phi_b = 0.95 \\ \phi_i = 0.80 \end{array}$ & $\begin{array}{c} r_b =0.13\,l \\ r_i=0.25\,l \end{array}$ & $\left(\begin{array}{cc} 0.064 & 0\\ 0 & 0.064 \end{array}\right)\,l^2$
  \end{tabular}
  \caption{Geometrical properties of the porous media considered
  in this work (see graphical representation in figure \ref{fig:geom}b).
  %Value of porosity $\phi$ is given.
  The subscript $i$ corresponds to the interface and the subscript $b$
  corresponds to the bulk. The porosity $\phi$ is defined as the ratio
  between the solid volume and the fluid volume.
  The last column shows the
  interior permeability tensor $\ten{K}$.}
  \label{tab:geom}
  \end{center}
\end{table}
%For comparison, we create effective model simulations corresponding to the same geometries. 
The averaged DNS will be compared to effective representations of the porous bed (figure~\ref{fig:geom}c). Within the
porous domain, we employ the Darcy's law, where the only unknown quantity 
is the pore pressure $p^-$. The interior permeability tensors~($\ten{K}$) for all
geometries are  listed in the last
column of table~\ref{tab:geom}. They were obtained by solving a set of Stokes
equations in a periodic unit cell in the bulk \citep{whitaker1998method,meibook}.
%At the interface with the free fluid we use the proposed boundary conditions (\ref{eq:poro-slip}, \ref{eq:poro-walln},\ref{eq:poro-pres-jump}).
A Neumann condition on pore pressure $\nabla p^- \cdot \vec{n}=0$ is enforced at solid boundaries of the cavity.
%, where $\vec{n}$ is the unit normal vector of the wall. 
This condition corresponds to zero fluid flux through the wall. Boundary conditions for the free fluid remain the same as in section~\ref{sec:valid-lid-rough}.
%
%Before we can obtain the model solution, we must choose  an appropriate interface location.
%must be chosen.
%The suggested interface location $z_i = 0.0$ for the considered porous materials are impractical due to meshing issues for the interface cell solver. Therefore
%
We place the interface at a distance $z_i = 0.1\,l$ above the solid structures;
the interface location $z_i = 0.0$ was not accessible due to meshing issues.
%which has been shown to yield very good agreement between the model and the DNS in the textured surface case (table~\ref{tab:lam-rough-itf-loc}). Next, 
We compute the effective
parameters appearing in boundary conditions (\ref{eq:text-slip},\ref{eq:poro-walln},\ref{eq:poro-pres-jump}) using
the procedure explained in 
sections~\ref{sec:text-slip},\ref{sec:text-walln} and \ref{sec:poro-pressure}.
% with the interface cell solver \citep{github2016UgisShervin}. 
%The obtained slip and transpiration lengths and the friction coefficients are reported in table~\ref{tab:poro-bc-params-results}. 
The slip and the transpiration
lengths, reported in
table~\ref{tab:poro-bc-params-results},  have nearly the same value for the three configurations because  the porous materials have similar porosity near the interface (table~\ref{tab:geom}).% and consequently at the same interface location the slip length and transpiration lengths are similar.

\begin{table}
  \begin{center}
\def~{\hphantom{0}}
  \begin{tabular}{p{11mm}p{12mm}p{12mm}p{16mm}p{16mm}p{16mm}p{10mm}p{10mm}p{10mm}}
    Config. & $L_{xx} / l$ & $M_{xx} / l$ & $f^{(1)}_x / \left( \mu / l \right) $ & $f^{(1)}_z / \left( \mu / l \right)$ & $f^{(2)}_x / \left( \mu / l \right)$ & $u_x / \bar{u}_x$ & $u^-_{z} / \bar{u}_z$ & $u_z / \bar{u}_z$  \\ \hline
       (i)   & $0.1516$ & $0.0856$ & $0.000$ & $-10.43$ & $\ \ \, 0.000$ & $0.958$ & $0.629$ & $1.127$ \\
       (ii)  & $0.1563$ & $0.0885$ & $2.125$ & $-7.948$ & $-1.541$ & $0.958$ & $0.661$ & $1.137$\\
       (iii) & $0.1538$ & $0.0866$ & $0.000$ & $-38.23$ & $\ \ \, 0.000$ & $0.986$ & $0.828$ & $1.153$
  \end{tabular}
  \caption{The slip length, the transpiration length, and the resistance
  coefficients $\vec{f}^{(1)}$ and $\vec{f}^{(2)}$ for
  the porous medium geometries shown graphically in figure \ref{fig:geom}(b).
  The last three columns show the ratio between the model and the DNS results for the slip
  and the transpiration velocities.}
  \label{tab:poro-bc-params-results}
  \end{center}
\end{table}

To validate the velocity conditions (\ref{eq:text-slip},\ref{eq:poro-walln}), we
sample the ensemble-averaged DNS and the effective model  at  
coordinates $(0.5\,H,z_i)$ and $(0.25\,H,z_i)$ for slip and transpiration
velocities, respectively. The ratio between the model predictions and the DNS
are given in last columns of table~\ref{tab:poro-bc-params-results}. It is clear that the slip velocity is predicted as
accurately as for the textured surfaces. 
In the second to last column of table~\ref{tab:poro-bc-params-results} we list the ratio
between the Darcy transpiration velocity $u_z^-$ -- sampled just below point $(0.25\,H,z_i)$ --
and the DNS result. The Darcy velocity alone is a rather inaccurate predictor and
has a relative error up to $37\%$. The agreement between
the TR model (\ref{eq:poro-walln}) -- that
augments the Darcy contribution with the
term containing the transpiration length -- and the DNS is better as the relative error
is smaller than $15\%$.
%

%To illustrate the importance
%of the proposed transpiration length (\ref{eq:poro-walln}), we also report the contribution of
%Darcy transpiration velocity $u_z^-$ -- sampled just below point $(0.25\,H,z_i)$ -- to the transpiration boundary condition (\ref{eq:poro-walln}).
%%For the transpiration velocity,
%The Darcy velocity alone is a rather inaccurate predictor and
%has a relative error up to $37\%$.
%The agreement between
%the TR model -- that augments the Darcy contribution with the term containing the transpiration length -- and the DNS is better as the relative error
%is smaller than $15\%$.

\begin{figure}
\centering
\subfloat[]{\includegraphics[trim = 0cm 0cm 0cm 0cm, clip, height=5.2cm]{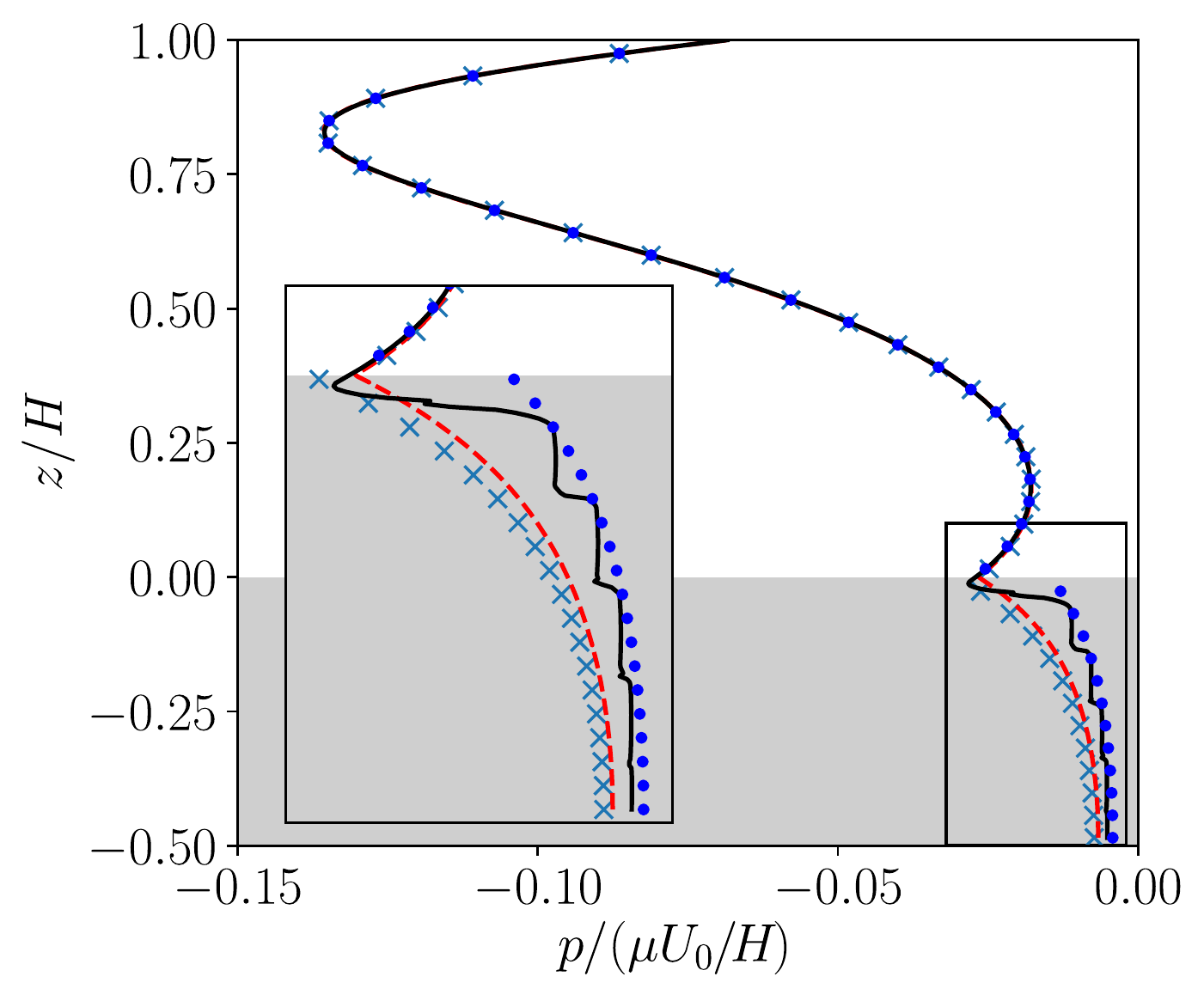}}
\hfill
\subfloat[]{\includegraphics[trim = 0cm 0cm 0cm 0cm, clip, height=5.2cm]{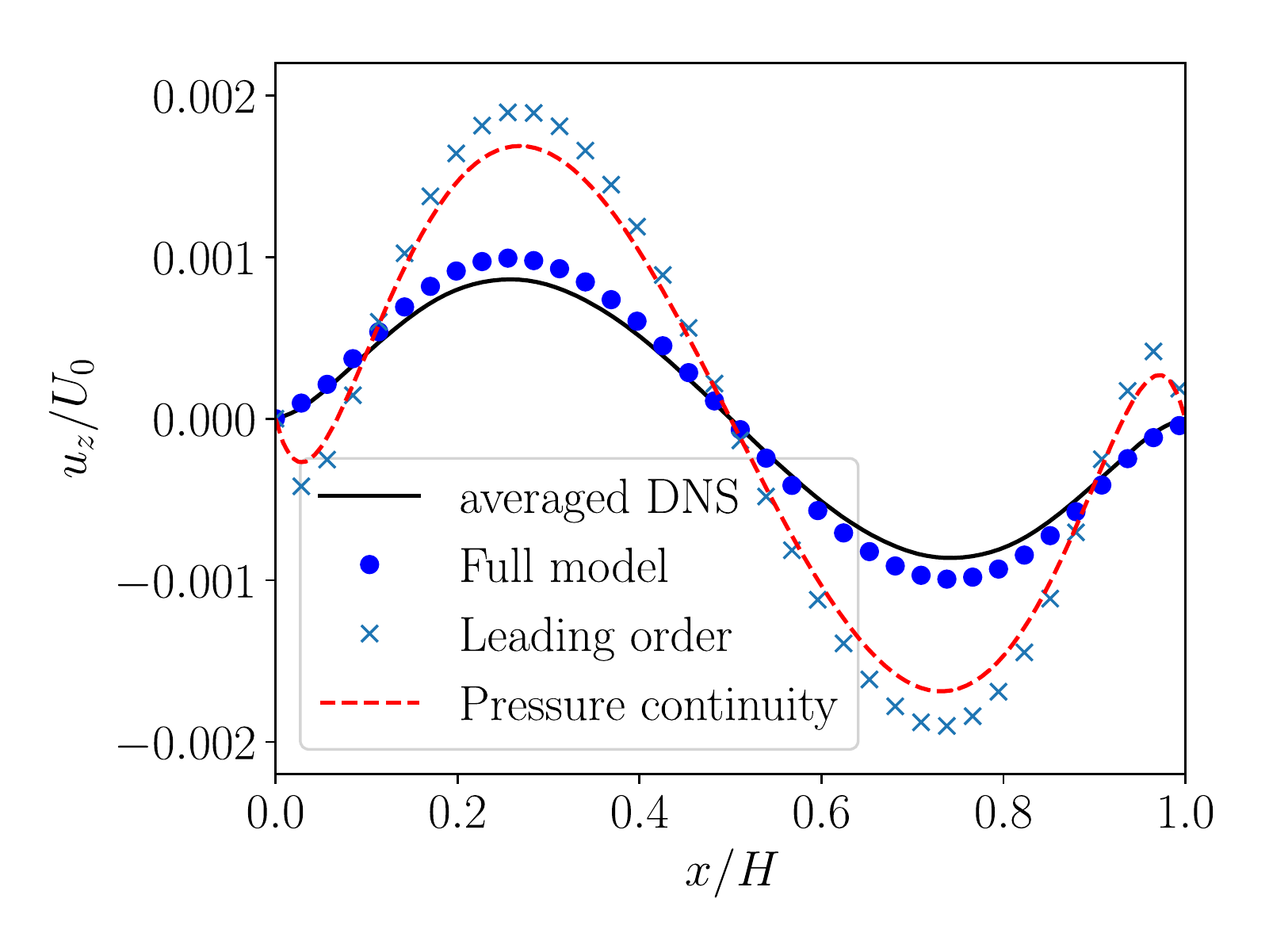}}
\caption{The pressure (a) and the transpiration velocity (b) profiles of the lid-driven cavity with layered isotropic porous bed (iii). Frame (a) shows
the distribution of pressure
along the vertical dashed line in figure~\ref{fig:geom}(a). 
The grey shaded region corresponds
to the porous material.
%The DNS results and the model results are shown concurrently. Interface location $z_i = 0.1\,l$.
}
\label{fig:ipor}
\end{figure}

To validate the pressure condition (\ref{eq:poro-pres-jump}), we analyze
the ensemble-averaged DNS of configuration (iii). 
The pressure $\hat{p}$ along the vertical dashed lined in figure \ref{fig:geom}(a)
is shown in figure~\ref{fig:ipor}(a)
with a solid black line. The region corresponding to porous domain is shadowed.
We observe that the pressure field undergoes a sharp variation when transitioning from the free fluid region to the porous medium, as shown by an inset in 
figure~\ref{fig:ipor}(a).
The sharp variation
indicates that there could be a pressure jump in the effective
representation.
%
%Note that this pressure jump would occur over a smaller and smaller
%macroscopic length if one would reduce the scale separation parameter
%(make the porous material more fine-grained). This would happen because
%there would be more and more porous material structures in the given
%distance and the pressure field
%distribution is correlated to the number of structures over which the field varies.
%
Note that there are microscale oscillations in the pressure field;
the employed ensemble average filters out only the microscale variations in the $x$
direction.
For comparison, the pressure obtained from the effective model 
is shown using dotted blue symbols in
figure~\ref{fig:ipor}(a).
We can observe that the agreement
between model predictions and DNS is good both in free fluid and
interior, while the sharp variation in the near vicinity of the interface
is not modelled. This is a direct consequence of having an infinitely
thin interface in the effective representation, which condenses all variations
near the boundary to a single line. 
%Nevertheless, the good agreement in free flow and porous regions proves that the proposed stress partitioning condition is accurately describing the main effect of this variation. 
The same quantitative agreement is observed for configurations (i) and (ii), but are not shown here. 
% as well, but the curves are similar
%and for brevity are not repeated.

To show the importance of the resistance coefficients $\vec{f}^{(1)}$ and $\vec{f}^{(2)}$, 
we carried out two more effective simulations. In the first one -- called ``leading order'' --
we set  $\vec{f}^{(1)} = 0$ in  the pressure condition  (\ref{eq:poro-pres-jump}).
This essentially corresponds to pressure
condition proposed
by \cite{marciniak2012effective,carraro2013pressure,carraro2018effective}.
In the second one -- called ``pressure continuity'' -- we
impose $p = p^-$, which has been a common
approach in the past \citep{ene1975,levy1975,hou1989,lacis2016}.
Results from the leading order model and the pressure continuity model
are reported in figure~\ref{fig:ipor}(a)
using crosses and a dashed curve, respectively. We observe that both
conditions result in a poor agreement between the model and the DNS
if compared to the TR model.

An inaccurate pressure condition can also influence the
flow field. To illustrate this,
we provide model results of interface
normal velocity in figure~\ref{fig:ipor}(b),
using the same symbols. We observe that the error in
pressure condition can lead to significantly different
-- and inaccurate -- 
vertical
velocity predictions. 
The coefficient 
$\vec{f}^{(1)}$ imposes larger resistance for the wall-normal velocity,
and, thus, it decreases the transpiration by precisely the correct amount.
%Although, as previously often observed,
%the vertical velocity is at least an order of magnitude smaller
%than tangential velocity, it
%is essential for capturing advective transport of nutritions,
%particles and ions between the free fluid and complex surface.  
%can be important -- as will be shown
%in the following section -- to correctly capture essential physical
%characteristics of flow surface interaction.
%Therefore the appropriate pressure condition leading to correct prediction of vertical velocity is important, if accurate velocity predictions at the interface are sought.

%\section{Turbulent flow over rough walls} \label{sec:turb}
\section{Role of the transpiration in a turbulent channel flow} \label{sec:turb}
%\section{Turbulent channel flow} \label{sec:turb}

\begin{figure}
\centering
\vspace*{10pt}
\subfloat[]{\raisebox{-0.5\height}{\includegraphics[width=0.49\linewidth]{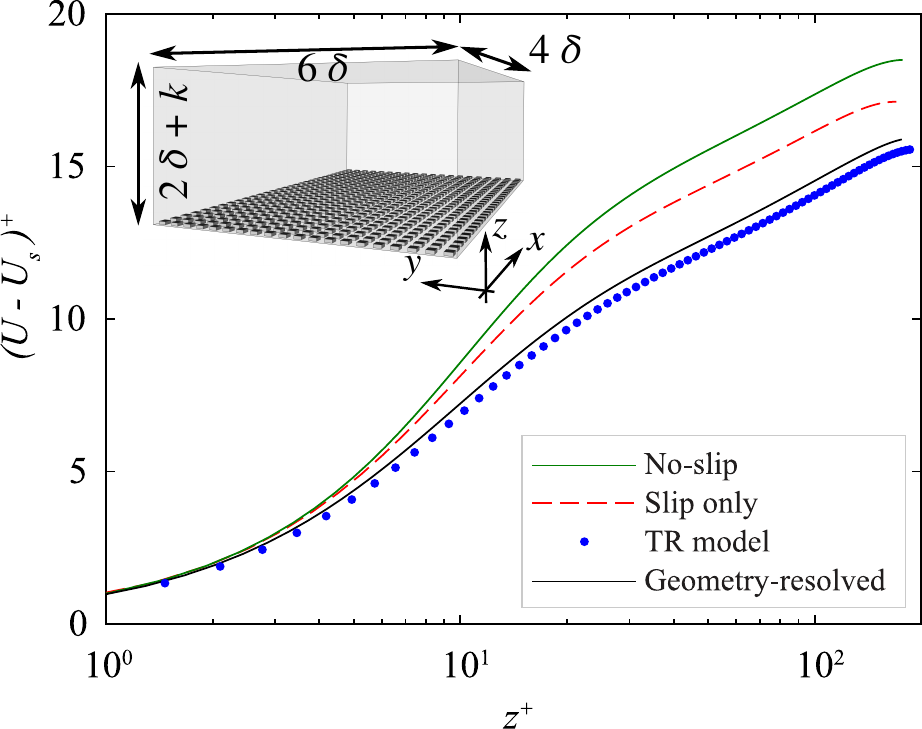}}}
\hfill
\subfloat[]{\raisebox{-0.5\height}{\includegraphics[width=0.49\linewidth]{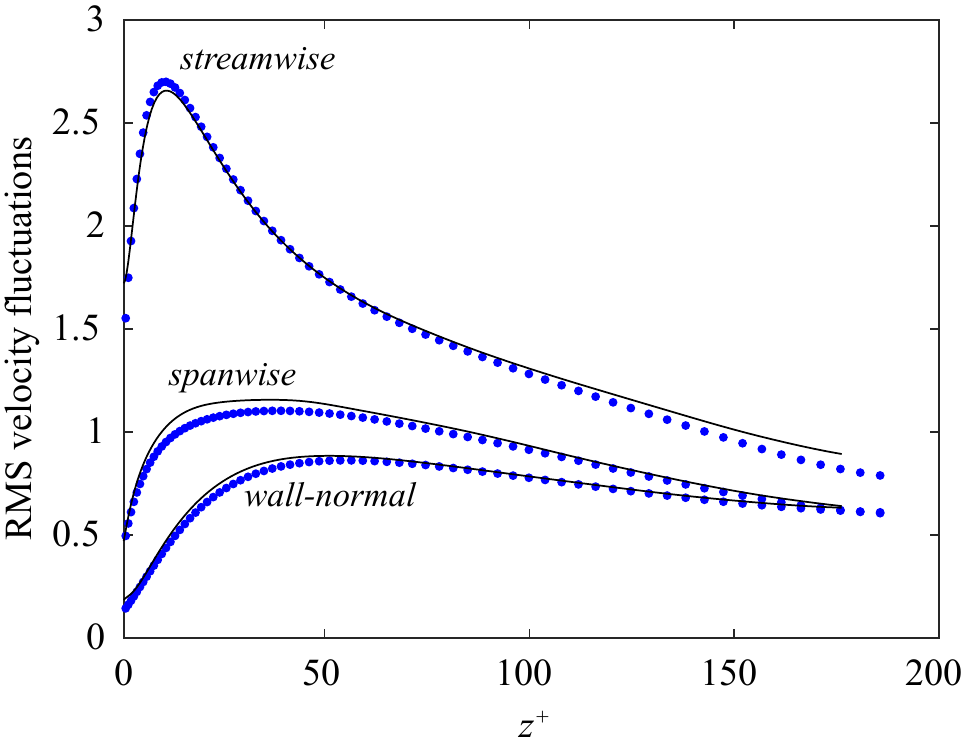}}}
%\subfloat[]{\raisebox{-0.5\height}{\includegraphics[width=0.49\linewidth]{figs/turbrmswithsmooth-roughSim.pdf}}}
%\subfloat[]{\raisebox{-0.5\height}{\includegraphics[width=0.49\linewidth]{figs/turbrmsall-roughSim.pdf}}}
%\subfloat[]{\raisebox{-0.5\height}{\includegraphics[width=0.49\linewidth]{figs/turbrmswithslip-roughSim.pdf}}}
%\vspace*{5pt}
\caption{Frame (a) shows the time and space averaged velocity profile (mean profile) for the turbulent channel flow. The inset in the frame (a) provides the simulation domain: $6\delta\times 4\delta\times (2+k)\delta $, where $k$ is
the height of the roughness elements and $\delta$ is the channel half-height. 
The dimensions of the periodic cuboid roughness elements are $0.1\delta\times
 0.1\delta\times k\delta$, with $k/\delta = 0.04$ and those are contained
in a periodic tile of $0.2\delta \times 0.2\delta$ along wall-parallel directions.
Frame (b) shows the root-mean-square of the velocity fluctuations for streamwise $u'_{rms}$, spanwise $v'_{rms}$ and wall-normal $w'_{rms}$ components. Here we compare results from geometry resolved simulations and three effective models.}
\label{fig:channel-rough}
\end{figure}

%In the previous sections, we have shown that the boundary conditions (\ref{eq:text-slip},\ref{eq:text-walln},\ref{eq:poro-walln}) are able to approximate the transfer of mass and momentum at the interface between a textured or porous surfaces and the free fluid for laminar flow conditions. 
%However, for these examples, the transpiration velocity has always been at least an order of magnitude smaller than the slip velocity (figure~\ref{fig:cavity-rough-result} and table~\ref{tab:lam-rough-itf-loc}). Therefore a natural question is, if it matters at all to capture the transpiration velocity correctly.
In this section, we demonstrate with a specific example -- fully-developed turbulent flow over
a textured surface -- that a small transpiration velocity can be crucial to capture the correct physics
of the problem. 
%  \citep{orlandi2006_3d}.
%
%We perform a turbulent flow simulation through a channel
Here, the domain consists of a periodic channel whose bottom surface
is covered with ordered cuboid roughness elements (see
%figure~\ref{fig:turbroughDNSdomain}.
inset in figure~\ref{fig:channel-rough}a). 

%The full domain size is $6\delta\times 4\delta\times (2+k)\delta $, where $k$ is the height of the roughness elements and $\delta$ is the channel half-height.  The dimensions of the periodic cuboid roughness elements are $0.1\delta\times  0.1\delta\times k\delta$, with $k/\delta = 0.04$ and those are contained in a periodic tile of $0.2\delta \times 0.2\delta$ along wall-parallel directions.
%

%To model this set-up using a homogenised approach, 
We define a flat interface on the crest plane of the cuboids. The region below this interface is  discarded in the
effective representation of the textured wall. We impose three different boundary conditions on the interface;
(i) no-slip condition corresponding to a smooth wall, (ii)
slip condition (\ref{eq:text-slip}) and (iii) the TR model,
including also the transpiration velocity (\ref{eq:text-walln}).
Since the cuboids
have the same geometry in both $x$ and $y$ directions and they are 
aligned with the chosen coordinate system, we have
%can immediately observe that 
$L_{xx} = L_{yy}$, $M_{xx} = M_{yy}$ and $L_{xy} = M_{xy} = 0$. 
The values of $L_{xx}$ and $M_{xx}$ are provided in table~\ref{tab:rough-turb}. They
were computed a priori 
by solving the fundamental problems (\ref{eq:itf-fund-1}--\ref{eq:itf-fund-2}) for a
cuboid roughness element in the interface unit cell using the procedure described
in section~\ref{sec:trm-text}.

%We have checked convergence using
%$\Delta x^+=4.14$, $\Delta y^+=2.45$, $\Delta z_w^+=0.3$, and $\Delta z_c^+=2.4$
%and observed that the change in measured time averaged stress at the wall
%was below $0.5\%$.

The  simulations 
%(details are provided few paragraphs below)
are carried out under conditions which lead
to $Re_{\tau} =u_\tau\delta/\nu \approx 180$, 
where $u_\tau$ is the friction velocity.
For all simulations, we impose a constant mass flux;
the driving pressure gradient is continuously adjusted. For more simulation
details, see appendix~\ref{app:turb-detail}.
Figure~\ref{fig:channel-rough}(a) shows the time and space averaged
mean velocity profiles for the three effective simulations in plus (or wall) units.
The averaged velocity on the
interface $U_s^+$ is subtracted from the mean flow $U^+$, such that all
profiles have zero value at the
crest plane of the roughness. 
% $u^+ - u^+_s$ We compute the time and space averaged mean velocity profile for all the simulations, shift all the profiles to have a zero value at the crest plane of the roughness and plot the results $u^+ - u^+_s$ in plus (or wall) units in figure~\ref{fig:channel-rough}(a).
%
% These units scale the mean velocity profile in the universal "log-law" defining a logarithmic profile in the defect layer as a function of the wall distance $z^+$ for smooth wall.
%
We note a downward shift of the logarithmic part of mean velocity profiles that increases from (i) to (iii).
The  logarithmic part of the mean flow can be represented by 
%For rough wall, this relation is still valid, but a constant shift of the defect layer
%is reported, see \cite{orlandi2006_3d}. The "log-law" writes
\begin{equation}
(U-U_s)^+=\frac{1}{\kappa}\ln\left(z^+\right)+B-\Delta U^+,
\label{eq:loglaw}
\end{equation}
where $\kappa = 0.392$ and $B = 4.48$ \citep{millikan1939critical,luchini2017}. Moreover,  $\Delta U^+$ is the roughness function that quantifies
the  shift in the mean velocity profile. 
%

%As a consequence of eq. (\ref{eq:loglaw}), the slop in the defect layer follows the von-Karman constant as illustrated in fig. \ref{fig:channel-rough}(a), and the shift between the smooth wall and the textured wall is clearly depicted in fig. \ref{fig:channel-rough}(a), see the difference between the green solid line and the model curves.
%We see that the slip simulations produce a certain shift of the mean velocity profile $\Delta U^+_s$ if compared to the no-slip simulation. Interestingly, adding a small transpiration velocity on top of the slip velocity produces an additional shift $\Delta U^+_2$ on top of the shift $\Delta U^+_s$, which is even larger in magnitude as the latter. This illustrates that this flow set-up is sensitive to the transpiration velocity and the presented TR model might give crucial insights in the flow physics. The importance of the vertical velocity for turbulent flows has been observed before \citep{jimenez2001,orlandi2006_3d,garcia2011}.

From figure~\ref{fig:channel-rough}(a), we observe that the slip velocity boundary condition produces a relatively small  shift ($\Delta U^+_s$)  compared to the simulations where both slip and transpiration velocity are imposed. The transpiration induce an additional shift $\Delta U^+_2$ which is in fact larger than $\Delta U^+_s$, despite that
the transpiration velocity is formally a higher-order boundary condition. 
%, which is even larger in magnitude as the latter. 
This illustrates the sensitivity of the turbulent channel flow  to  the transpiration velocity as
%and the presented TR model might give crucial insights in the flow physics.  The importance of the vertical velocity for turbulent flows has been observed before \citep{jimenez2001,orlandi2006_3d,garcia2011}.
recognised in earlier studies \citep{jimenez2001,orlandi2006_3d,garcia2011}.

%To arrive with a more quantitative conclusions, we set-up a fully resolved simulation.
We also carried out DNS using an immersed boundary method  to resolve the
flow around cuboids \citep{breugem2006}. 
%For all these turbulence simulations, we impose a constant mass flux; the driving pressure gradient is continuously adjusted.
%
%\begin{figure}
%\centering
%\includegraphics[width=0.42\linewidth]{figs/Turb_DNS_domain_size}
%\caption{The rough bottom channel wall and the dimensions of resolved simulation domain.\label{fig:turbroughDNSdomain}}
%\end{figure}
%
%
Figure~\ref{fig:channel-rough}(a) shows the time and space averaged mean flow of the geometrically resolved DNS.
In order to make a comparison with effective simulations, we  subtracted the mean slip velocity at the crest plane for the DNS as well.
%
% the obtained mean profile according to the slip velocity at the interface and plot the DNS results along with the effective model results in figure~\ref{fig:channel-rough}(a). 
%
We observe that there is a relatively good agreement with the mean velocity profile of the TR model.
%, while for the slip boundary alone there is a larger discrepancy.
%Thus, the TR model significantly improved the shift of the mean velocity induced by the roughness compared to the slip boundary condition. 
A good agreement is also observed for the rms velocity fluctuations (figure \ref{fig:channel-rough}b).

%The TR model is, thus, able to modify the near wall turbulence in a correct way, in contrast to conditions based on slippage only. 

%induce the increased isotropy of the turbulence keeping the turbulent streaks as well as increasing the rms fluctuations along the spanwise and the wall-normal direction as observed in the geometry-resolved DNS.

\begin{table}
\begin{center}
\def~{\hphantom{0}}
\begin{tabular}{p{35mm}p{12mm}p{12mm}p{12mm}p{12mm}}
 & $k/\delta$  & $L_{xx}/\delta$ & $M_{xx}/\delta$ & $Re_{\tau}$ \\[5pt]
No-slip          & -        &  -  & - & 178.76 \\[5pt]
Slip only  & 0.04   & 0.01146 & 0   & 172.32 \\[5pt]
TR model & 0.04   & 0.01146 & 0.01602 & 188.02 \\[5pt]
Geometry-resolved  & 0.04        &  -  & - & 184.70 \\
\end{tabular}
\caption{Friction Reynolds numbers ($Re_{\tau}$) at the bottom wall of the turbulent channel flow with smooth, rough and slip boundary conditions at the bottom wall. Here $k$ and $\delta$ represent roughness elements' height and channel half height, respectively. $L_{xx}$ and $M_{xx}$ are coefficients used in equations (\ref{eq:text-slip},\ref{eq:text-walln}). For the considered roughness geometry $L_{xx}=L_{yy}$, $M_{xx}=M_{yy}$ and $L_{xy} = L_{yx} = M_{xy} = 0$.}
%{\color{red} I checked that at $z^+ = 0.36$ we get
%$u_x / \bar{u}_x = 1.023$ for the Rough\_0.04
%case if compared to DNS simulations. Not sure if we should include this
%in the paper.}}
\label{tab:rough-turb}
\end{center}
\end{table}

%\begin{table}
%\begin{center}
%\def~{\hphantom{0}}
%\begin{tabular}{p{45mm}p{12mm}p{12mm}p{12mm}p{12mm}}
% & $k/\delta$  & $L_{xx}/\delta$ & $M_{xx}/\delta$ & $Re_{\tau}$ \\[5pt]
%Smooth DNS          & -        &  -  & - & 178.76 \\[5pt]
%Geometry-resolved DNS    & 0.04        &  -  & - & 184.70 \\[5pt]
%TR model ''With transpiration'' & 0.04   & 0.01146 & 0.01602 & 188.02 \\[5pt]
%TR model ''No transpiration''   & 0.04   & 0.01146 & 0   & 172.32 \\
%\end{tabular}
%\caption{Friction Reynolds numbers ($Re_{\tau}$) at the bottom wall of the turbulent channel flow with smooth, rough and slip boundary conditions at the bottom wall. Here $k$ and $\delta$ represent roughness elements' height and channel half height, respectively. $L_{xx}$ and $M_{xx}$ are coefficients used in equations (\ref{eq:text-slip}--\ref{eq:text-walln}). For the considered roughness geometry $L_{xx}=L_{yy}$, $M_{xx}=M_{yy}$ and $L_{xy} = L_{yx} = M_{xy} = 0$.}
%%{\color{red} I checked that at $z^+ = 0.36$ we get
%%$u_x / \bar{u}_x = 1.023$ for the Rough\_0.04
%%case if compared to DNS simulations. Not sure if we should include this
%%in the paper.}}
%\label{tab:rough-turb}
%\end{center}
%\end{table}

%The relaxation of the no-penetration condition can be appreciated by comparing $Re_{\tau}$ for all simulations given in table~\ref{tab:rough-turb}. 
The friction Reynolds number $Re_{\tau}$ for all simulations is given in table~\ref{tab:rough-turb}.  For channel flow with rectangular cuboid roughness elements, the friction velocity (and thus the skin-friction drag) at the rough wall is larger compared to that of a smooth wall~\citep{orlandi2006_3d}. However, effective simulations without the transpiration predict a reduction in the skin-friction drag: this can be observed by a smaller $Re_{\tau}$ for the simulations denoted as ``Slip only''.  
In contrast, the TR model is able to modify the near wall turbulence in a correct way, predicting the roughness-induced drag increase.
%Such drag modification behaviour with only slip boundary conditions is already well documented by~\cite{busse2012}, who reports that the streamwise slip induces drag reduction while the spanwise slip induces drag increase. 

%The $Re_{\tau}$ is computed using 
%To determine the friction velocity $u_\tau = \sqrt{\tau_w / \rho}$, 
Additional insight into the role of the transpiration can be gained from the total wall stress, given by
\begin{equation}
\tau_w=\underbrace{\mu\frac{\partial U}{\partial z}}_{\tau_V}-\underbrace{\rho \overline{u_x'u_z'}}_{\tau_R}, \label{eq:turb-w-stress}
\end{equation}
where $\rho$  is the fluid density. Here, the over-bar denotes time and space averaged quantities, and $(\bm{\cdot})'$ represent turbulent fluctuating quantities. The first term ($\tau_V$) is the viscous stress, and the second term ($\tau_R$) is the Reynolds stress. %, which quantifies momentum transport due to turbulent fluctuations. 
For channel flow with slippage only, the wall normal fluctuations are zero ($u_z'=0$) at the wall. Hence, equation (\ref{eq:turb-w-stress}) simplifies  to $\tau_w = \tau_V$ and only the viscous stress is modified by the boundary condition.  In contrast, when also the transpiration condition is imposed, $u_z'\neq 0$, which allows for a direct modification of the Reynolds stress at the wall. More thorough analysis
of transpiration velocity effect on the turbulence
can be found in papers by~\cite{gomez2018manipulation,garcia2019control}.

% shows an increase of the $Re_{\tau}$ value, which is consistent with the geometry-resolved simulation. A direct ascertainment is $\tau_R \neq 0$, modifying the skin friction drag due to wall normal fluctuations. However, the mechanism behind this modification includes deeper considerations and we refer to \cite{gomez2018manipulation} for a better understanding. In addition, the TR model predicts a larger value of the friction Reynolds number compare to the geometry-resolved simulation. This discrepancy is due to a slightly over prediction of the turbulent Reynolds stress of our model.

%%% FOR NEXT PAPER: The sensitivity to transpiration velocity has already been used to control turbulent flow, \cite{lumley1998} to delay turbulence transition or to enhance flow mixing for instance. Although slip and transpiration conditions have been intensively investigated in turbulent flow, boundary conditions that model consistently turbulent flows over textured surfaces remains not obvious. Our investigations on the turbulent channel flow shows, first, that the transpiration velocity is again crucial in dictating the overall dynamics of rough wall turbulent flows, and, second, that the TR model leads to an appropriate boundary condition connecting the slip and transpiration velocity to substitute textured surface in turbulent flows. 

%It should be noticed that the TR model has some limitations to predict turbulent flow over rough walls. 
The TR model is based on the creeping flow or linearity assumption (\ref{a1}).
Naturally the
model is expected to work, if the roughness size is below the size of the viscous
sub-layer. The roughness considered here is so-called transitional
roughness \citep{jimenez2004turbulent}, which is slightly larger than the
viscous sub-layer ($k^+ \approx 7$).
The model, however, still provides a reasonable approximation of the DNS,
despite the fact that there are inertial effects present in the flow between textured elements.
For even larger
roughness elements, 
%that are higher or wider than the characteristic near-wall turbulent structures,
inertial effects inside the textured surface will become more important,
rendering the 
fundamental problems (\ref{eq:itf-fund-1}--\ref{eq:itf-fund-5}) inaccurate.
From our experience, the TR model will fail for
roughness heights around $k^+ \approx 20$.

An empirical model for describing substantially larger surface textures
would be required to capture non-linear effects, such as sweeps and ejections in the boundary
layer \citep{breugem2006} and transpiration velocity due to the pressure
fluctuations \citep{garcia2019control}. This is however out in scope of this
work.% and is left for the future development.

%========================================================================================

\section{Comparison to the multi-scale expansion} \label{sec:discuss}

We compare the effective conditions in the TR
model (\ref{eq:text-slip},\ref{eq:text-walln},\ref{eq:poro-walln},\ref{eq:poro-pres-jump}) with a set of conditions
obtained from a multi-scale expansion (MSE).
%Through the examples of the TR model for both textured and porous surfaces, we have noticed that the correction obtained in the transpiration velocity is always at least an order of magnitude smaller compared to the slip velocity. This naturally raises few questions such as (i) what orders of different terms of the TR model and (ii) are presented terms sufficient for practical purposes? To give an answer to these questions,  We carried out a formal multi-scale expansion (MSE) to obtain a mathematically well founded and complete set of boundary conditions. 
Appendix~\ref{app:mse} provides the essential components of the derivation. A more in depth analysis and derivations
of the MSE model will be presented in a separate paper.
Here, we focus our attention on two aspects; first, one-to-one comparison between the terms of the TR model and the MSE and, second, the accuracy of the TR model compared to the MSE. 

\subsection{One-to-one comparison between the TR model and the MSE}
An expansion up to the order $\ordest{\ord}$ results in the following boundary conditions (\ref{appeq:bc-mse-vel}) for the free fluid velocity at the interface,
\begin{equation}
\vec{u} = \underbrace{\ten{L}_e \cdot \frac{\vec{\tau}_e}{\mu}}_{\ordest{U^s}}
- \underbrace{\ten{K}_e \cdot \frac{\nabla p^-}{\mu}
+ \ten{M}_{e} : \frac{\nabla\vec{\tau}_e}{\mu}}_{\ordest{\ord U^s}}
+ \ordest{\ord^2\,U^s}, \label{eq:vel-bc-from-mse}
\end{equation}
where $\vec{\tau}_e = \mu \left(\pd_z u_x + \pd_x u_z, \pd_z u_y + \pd_y u_z, 
2 \pd_z u_z \right)$ is the fluid shear stress 
(containing the symmetric part) of the free fluid 
and $U^s$ is a characteristic magnitude of the slip velocity.
The tensor $\ten{M}_e$ is a 3rd-rank tensor with $81$ elements, 
while the tensors $\ten{L}_e$ and $\ten{K}_e$ are both 2nd-rank tensors with
$9$ elements.
%It turns out that the last row and column of the $3 \times 3$ tensor $\ten{L}_e$
%contains only zeros.
The double dot operation between  a 3rd-rank tensor $\ten{A}$ and a 2nd-rank tensor $\ten{B}$ is defined
as $\ten{A} : \ten{B} = A_{ijk} B_{jk}$, where summation over repeating indices is
implied. 

The leading order term ${\ordest{U^s}}$ of the
velocity condition \eqref{eq:vel-bc-from-mse} is the slip term.
%and  two higher order terms (${\ordest{\ord^2 U^s}}$).
%With brackets below the expression we illustrate the different orders of terms.
%The slip term in this boundary condition to the slip boundary condition in the TR model (\ref{eq:text-slip}) we observe that the 
The slip tensor $\ten{L}$  in the TR model (\ref{eq:text-slip})
corresponds to the upper left $2 \times 2$ block of $\ten{L}_e$.
In other words, the TR model contains the leading order MSE term with a reduced
expression of the shear stress tensor. %Other terms for the slip velocity are completely neglected.
From mass conservation arguments, it can be shown that the last row and column of tensor $\ten{L}_e$ appearing in \eqref{eq:vel-bc-from-mse} are zero.  Consequently there is no transpiration velocity at ${\ordest{U^s}}$.
%
%This explains why in all the validations examples the transpiration velocity was at least an order of magnitude smaller than the slip velocity.
%

There are two higher order ${\ordest{\ord U^s}}$ terms in the velocity condition \eqref{eq:vel-bc-from-mse}; a Darcian term related to the pore pressure gradient and a term related to the variation of the shear stress.
In the TR model, the Darcy contribution to the tangential velocity components at the interface is neglected. In other words, the TR model has the Darcy contribution only for the wall-normal transpiration component (\ref{eq:poro-walln}).
One may again show from mass conservation, that the last row of $\ten{K}_e$ in 
equation \eqref{eq:vel-bc-from-mse}
%Another interesting observation is that the part of the tensor $\ten{K}_e$ (last row) 
is equal to the last row of the interior permeability tensor $\ten{K}$. Consequently,
the Darcy term 
$u_z^-$ in the TR model (\ref{eq:poro-walln-Darcy}) corresponds to $(\ten{K}_e \cdot \nabla p^-/\mu)\cdot \hat n$.

Finally, the term related to the variation of the shear stress in
%Consequently, that term in the transpiration velocity is exactly the interior Darcy velocity. 
equation (\ref{eq:vel-bc-from-mse}) is compared with the transpiration boundary conditions in the TR model (\ref{eq:text-walln}). We observe that the TR model contains some of the next order terms arising from the variation of the shear stress; the TR model contains only $2$nd-rank tensor corresponding to tangential shear stress variations in the tangential directions, which is in contrast
to the
full $3$rd-rank tensor in the MSE corresponding to all shear stress variations in all directions.

The boundary condition for the pressure derived using  MSE (\ref{appeq:bc-mse-p}) is
\begin{equation}
p^- - p = \underbrace{\vec{b} \cdot \vec{\tau}_e}_{\ordest{\Delta P}}
- \underbrace{\vec{a} \cdot \nabla p^-
+ \ten{C} : \nabla \vec{\tau}_e}_{\ordest{\ord \Delta P}}
+ \ordest{\ord^2\,\Delta P}, \label{eq:p-bc-from-mse}
\end{equation}
where $\Delta P$ is a characteristic magnitude of a pressure drop in the system.
Here, $\vec{b}$ and $\vec{a}$ are vectors and $\ten{C}$ is a 2nd-rank
tensor.% with size $3 \times 3$. 

The leading term ${\ordest{\Delta P}}$ of expression \eqref{eq:p-bc-from-mse} induces a pressure jump proportional to the shear stress. 
%
%Again, terms appearing at different orders are indicated. 
It can be shown through mass conservation and force balance that the last element
in $\vec{b}$ (corresponding to shear stress $2 \mu\, \pd_z u_z$) is always equal to $-1$. Therefore this term can be transferred to the left hand
side and grouped together with ``$-p$'' to yield the total free fluid stress, as
appearing in the TR model (\ref{eq:poro-pres-jump}). 
%
%Comparing the MSE expression (\ref{eq:p-bc-from-mse}) with the pressure condition in the TR model (\ref{eq:poro-pres-jump}) 
Further, we assert that the first two elements of the vector $\vec{b}$ in expression (\ref{eq:p-bc-from-mse}) corresponds to the friction factor $\vec{f}^{(2)}$ in equation (\ref{eq:poro-pres-jump}). This can be confirmed 
by replacing the shear stress in expression \eqref{eq:p-bc-from-mse} with $\ten{L}^{-1}\vec{u}_t$. Consequently there is a full overlap of the leading order pressure jump terms between the TR and the MSE.
%
%Comparing the MSE expression (\ref{eq:p-bc-from-mse}) with the pressure condition in the TR model (\ref{eq:poro-pres-jump}) further, we observe that the  first two elements of the vector $\vec{b}$ corresponds to the friction factor $\vec{f}^{(2)}$, 
%
%
The two higher order ${\ordest{\ord \Delta P}}$ terms in the pressure condition \eqref{eq:p-bc-from-mse} are considered next. The vector $\vec{a}$ corresponds to the friction factor $\vec{f}^{(1)}$, which is confirmed 
by  replacing the pore pressure in equation \eqref{eq:p-bc-from-mse} with $\ten{K}^{-1}\vec{u}^-$.
%the help of inverse slip length and slip velocity and the pore pressure gradient with the help of inverse permeability and Darcy velocity. 
%% I do not understand this::::: Note that this transform is possible only within the terms selected for the TR model,  because it is not possible to take a interface normal derivative of a boundary condition defined only on the interface. 
Finally, the term corresponding to variations of the shear stress is completely neglected for the pressure condition in the TR model.

\begin{table}
  \begin{center}
\def~{\hphantom{0}}
  \begin{tabular}{p{20mm}|p{30mm}p{40mm}p{25mm}cccc}
   & Tangential velocity & Normal velocity & Pressure jump \\
        & $\vec{u}_{t}$ & $u_n$ & $p^- -p$ \\\hline\\[5pt]
       Leading order & $\mu^{-1}\ten{L} \cdot \vec{\tau}$ & $0$ &
       $- 2 \mu\,  \pd_z u_z + \vec{f}^{(2)} \cdot \vec{u}_t$ \\[0.5cm]
       Next order & $ 0 $ & $u_n^- - \ten{M} : \nabla_2 \vec{u}_t$  & $\vec{f}^{(1)} \cdot \vec{u}^-$\\[0.5cm]
  \end{tabular}
  \caption{Summary of the TR model boundary condition terms at the orders, at which
  corresponding terms emerge from the formal multi-scale expansion.}
  \label{tab:order-summary}
  \end{center}
\end{table}

In table~\ref{tab:order-summary} we group the different
terms of the TR model according
to the order at which the corresponding terms emerge in the multi-scale expansion. 
The slip velocity in the TR model contains only the leading order term with
a reduced shear stress vector, while 
the transpiration condition and the pressure condition contains all the leading
order contributions and some of next order corrections.
%This answers question (i) and provides an overview of the orders of the different boundary condition terms.

\subsection{Accuracy of the TR model compared to the MSE}
%To determine the accuracy of TR boundary conditions, 
We compare the flow and the pressure in the lid-driven cavity with a textured and a porous surface computed from fully resolved ensemble averaged
DNS with three different effective models; 
(i) the zeroth-order model, containing only leading order terms
from the MSE condition (\ref{eq:vel-bc-from-mse}--\ref{eq:p-bc-from-mse}), (ii) the first-order
model, containing all terms from the MSE
condition (\ref{eq:vel-bc-from-mse}--\ref{eq:p-bc-from-mse}) and (iii) the TR
model.

\begin{figure}
\centering
\subfloat[]{\includegraphics[height=4.5cm]{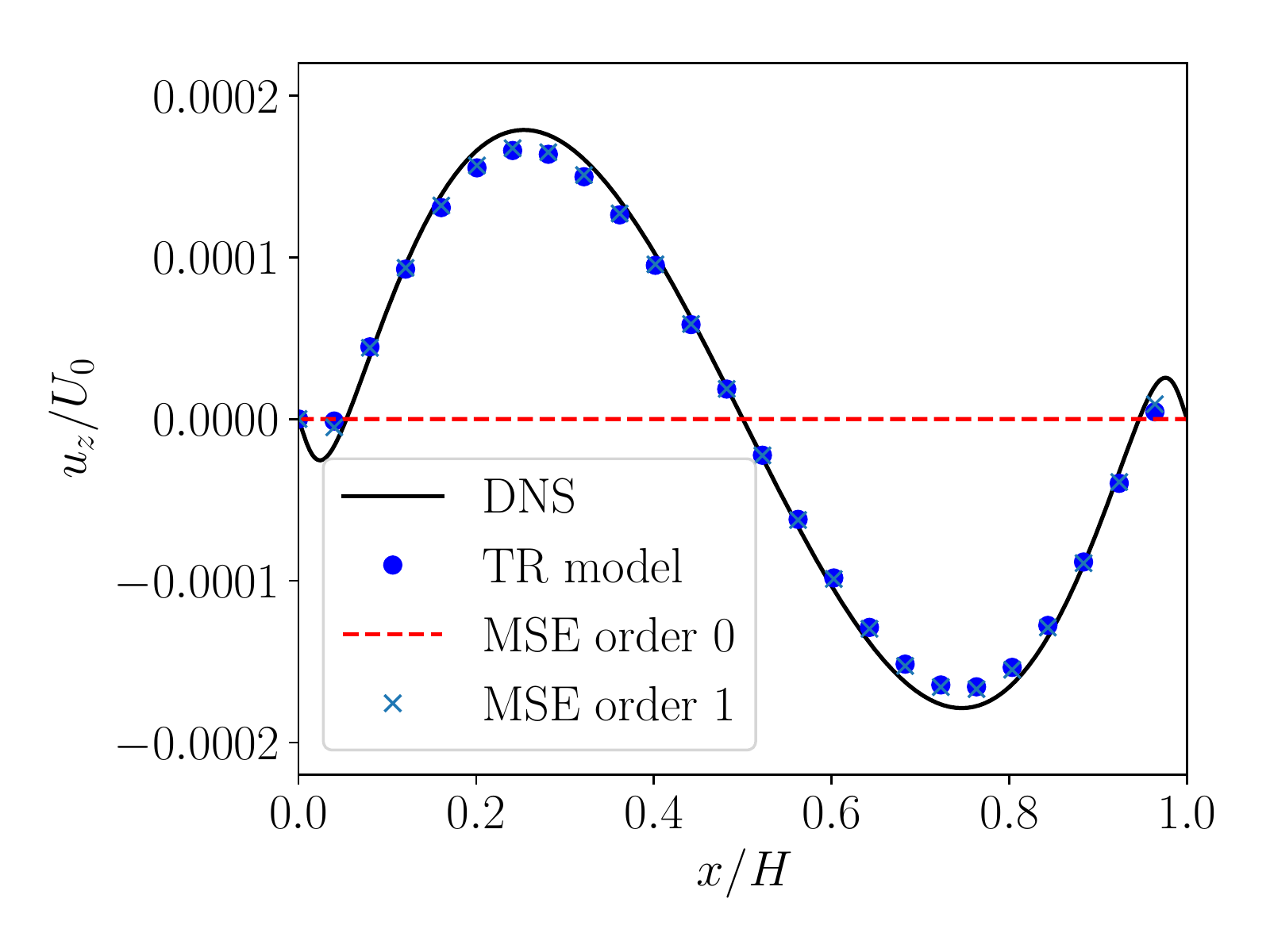}}
\subfloat[]{\includegraphics[height=4.5cm]{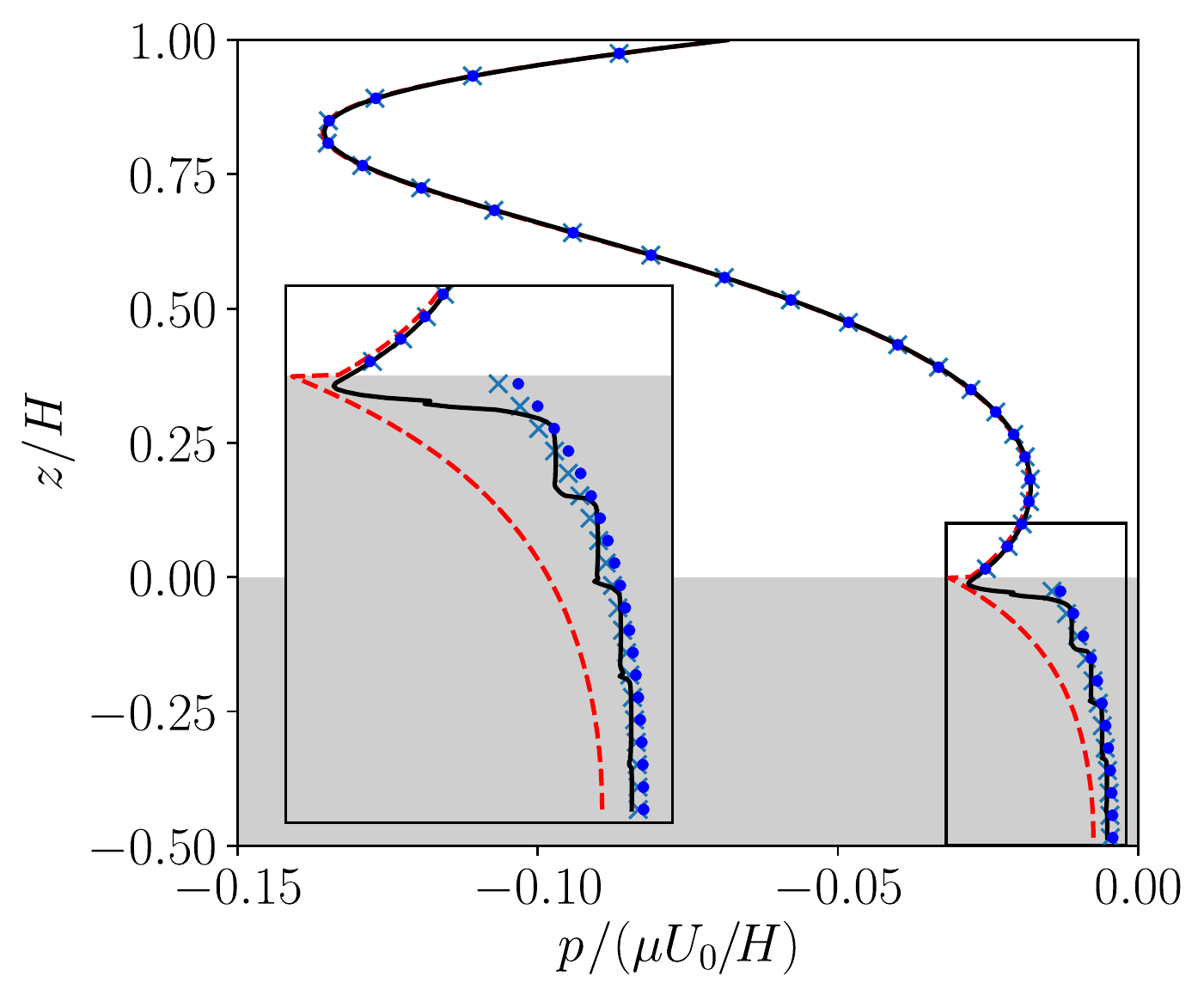}}
\caption{(Color online.) The transpiration velocity along
the interface between the flow in the lid-driven cavity and
the rough surface (a). Pressure distribution along the vertical slice for
the cavity flow over the layered porous surface (b). Interface location $z_i = 0.1\,l$.}
\label{appfig:model-comp-empirical}
\end{figure}

The transpiration velocity along the interface -- at the same
textured wall as discussed in section~\ref{sec:valid-lid-rough} -- is shown 
in figure~\ref{appfig:model-comp-empirical}(a). The pressure distribution along the
vertical slice -- of the same layered geometry from section~\ref{sec:valid-lid-poro} -- 
is shown in figure~\ref{appfig:model-comp-empirical}(b).
The interface in both cases is located at $z_i = 0.1\,l$. 
%
%During those tests we have observed that the TR model provides a good approximation of the transpiration velocity as well as the pressure jump. 
As expected, the TR model has a clear improvement over the zeroth-order MSE model.
It is observed that the first-order 
 MSE model
and the TR model nearly overlap for both the transpiration and the pressure fields. 
%
%In addition, the choice to neglect certain terms from MSE allows for a reduced set of fundamental problems to be solved and all the framework can be explained in mathematically less involved fashion. We believe that this is an important advantage of the model; 
We can thus conclude that the TR model provides nearly as good approximation as the MSE model, but with significantly reduced complexity. % of the TR model is significantly reduced if compared to the full formal MSE approach. 
%We believe that his answers the question (ii) and shows that the TR model is sufficient for practical purposes.

As a final remark, we note that  the TR model is not mathematically (or asymptotically)
fully consistent. If mathematical rigour is sought, all the next order terms for the slip velocity,
transpiration velocity and pressure condition should be taken into account. 
However, as we have demonstrated using the turbulent channel flow, a small
transpiration velocity gives a similar magnitude of shift in the mean velocity
profile as the large slip velocity (figure~\ref{fig:channel-rough}a).
Consequently, small corrections of slip velocity would not change results significantly,
while the introduction of the small transpiration velocity leads to notable modification of the mean flow.
In other words, it can be argued that -- for certain physical problems --  some higher-order terms are more important than  others. 
%Thus the TR model should be viewed as an engineering model capable of describing problems, which are sensitive to the small transpiration velocity.

\section{Conclusions} \label{sec:concl}

%This characterisation tool
%The TR model developed in this paper is useful in a number of ways. First, 
The TR model provides a set of accurate effective boundary conditions suitable for modelling free-fluid interacting with  rough and porous surfaces. % using numerical investigations. 
%, which have been inspired from a formal multi-scale expansion.
%These conditions are suitable for modelling free fluid interaction with  rough and porous surfaces.
These boundary conditions can be incorporated into computational fluid dynamics codes and thus enable investigations of how fully anisotropic textures and porous materials interact with external fluids. We have validated the TR model for creeping flows over textured and porous surfaces. Moreover, based on our investigations, we suggest to place the interface as close to the solid structures as possible without intersecting the solid structures.

The values of the coefficients within the TR model -- the slip length tensor $\ten{L}$, the transpiration length tensor $\ten{M}$, the resistance vectors $\vec{f}^{(1)}$ and $\vec{f}^{(2)}$ -- provide direct information of transfer of mass and momentum that can be expected when the surface interacts with a flow.
These coefficients can be computed for any surface topology using five fundamental
Stokes problems and are properties only of the surface itself.
To understand the significance of this, we make an analogy with a bulk porous material. In this case, the permeability is an established measure that characterises the ability of the porous material to transmit fluids. This measure is invaluable in understanding and designing porous materials for applications. In a similar way, we believe that slip, transpiration and resistance coefficients have a physical meaning on their own.
%
%% (I dont think we need this....)Note that in the thick surface case (such as porous media), this set of parameters has to be complemented with the surface specific characteristics, specifically interior permeability $\ten{K}$. This measure also influence how the surface interacts with the overlaying fluid (through the interface normal Darcy velocity).

The TR model is derived by making three assumptions; (i) a creeping flow near the surface texture; (ii) scale separation between the texture size and the flow length scale and; (iii) a repeating surface geometry. This means that the proposed model is a type of homogenised interface condition. %, as often derived using different homogenisation approaches. 
By using a formal multi-scale expansion (MSE), we have identified the theoretical orders of different terms present in the TR model. We have also shown how the TR model and
MSE model predictions compare using the lid-driven cavity as a test bed. Based on this comparison, we have justified that certain higher-order terms of the MSE model can be neglected. This results in a set of effective boundary conditions that are much simpler to implement and use compared to the full MSE conditions. %, without sacrificing much pf the physical 
%slmpler by observation that the TR model is very close to the full MSE model. 
%

For configurations where there exists an intrinsic hydrodynamic sensitivity to wall-normal velocity, the transpiration velocity may become as important as the slip velocity, although the former is, from an asymptotic viewpoint, a higher order correction. We have shown one such example here, namely the turbulent channel flow, for which the friction at the rough wall has a direct contribution from wall-normal velocity fluctuations. Similarly, the resistance terms in the normal stress balance condition for porous media can be shown to have different sizes via scaling analysis, but the relevance of the terms can only be determined  when the targeted application is taken into consideration. Here, we have shown that a so-called layered porous material need a higher-order resistance coefficient in order to physically capture the ``layering effect''. In nature, there is an abundance of porous materials with inhomogeneous layers; one example is the otolith structure inside human ear, which is part of our vestibular apparatus. Otoliths (calcium carbonate crystals) are located on top of a gel membrane, in which hairy sensory structure is located. The transfer of external fluid into these types of complex materials thus requires the higher order description based on transpiration length and resistance coefficients.

The generalisation of the TR model to elastic and poroelastic surfaces is relatively straightforward by applying the model locally, fixed to the displacing solid \citep{lacis2019noteTRelast}.
%One has to formulate the boundary conditions presented here in
%the local Lagrange coordinate 
Furthermore, the TR model can also be used for curved interfaces
using a coordinate transformation, provided that the curvature of the interface
is larger than characteristic surface length $l$.
Another interesting direction is the extension of the TR model by
considering the slip length, the transpiration length and the resistance coefficients
as spatially varying, time dependent
% presented in this paper can be extended to become spatially varying, time dependent
as well as flow dependent (for example, shear or Reynolds number dependent).
%These coefficients can also be extended to curved interfaces, provided that curvature is much larger than the intrinsic length of the surface.
Indeed, extensions of this kind may open up exciting modelling opportunities in applied problems including turbulent flows, heat transfer,
nutrition transport, etc.

\section*{Acknowledgements}
U.L. and S.B. acknowledge funding from Swedish Research Council
(INTERFACE center and grant nr. VR-2014-5680).
S.Y. acknowledges funding from the European Union's Horizon 2020
research and innovation programme under the Marie Sklodowska-Curie grant
agreement number 708281.
%The turbulent flow
S.P. acknowledges funding from the Swiss National Science foundation 
(project nr. P2ELP2\_181788).
Numerical simulations were performed on resources provided by the Swedish National Infrastructure for Computing (SNIC) at PDC and HPC2N.
Authors acknowledge Prof. Luca Brandt and Dr. Marco E. Rosti for
sharing their turbulent code and assisting with implementation of boundary
conditions. Authors thank Prof. Alessandro Bottaro for his valuable comments
about computing the transpiration length.
% and three anonymous reviewers for their time spent on scrutinising our work.

\appendix

\section{Multi-scale analysis for boundary conditions} \label{app:mse}

In this appendix, we provide a derivation of the boundary condition terms
at different orders using multi-scale expansion (MSE). The derivation follows
the approach previously used by \cite{lacis2016}.

\subsection{Dimensionless Navier-Stokes equations} \label{app:mse-nondim}

The starting point is incompressible dimensional Navier-Stokes (NS) equations, defined
in all space filled by fluid. These equations
are rendered dimensionless using relationships
\begin{equation}
\hat{\vec{u}} = U^s\, \hat{\vec{u}}',
\qquad \hat{p} = \Delta P\, \hat{p}',
\qquad \vec{x} = l\, \vec{x}',
\qquad t = t_s \, t', 
\label{appeq:non-dim-eq}
\end{equation}
where primed variables are dimensionless. Here, $U^s$ is characteristic slip velocity
near the surface, $\Delta P$ is characteristic pressure drop in the system and $t_s$
is characteristic time scale. Using expression (\ref{appeq:non-dim-eq}), the
dimensionless Navier-Stokes equations read
\begin{align}
Re \left[ St\,\pdt \hat{\vec{u}}' + \left( \hat{\vec{u}}' \cdot \nabla \right)
\hat{\vec{u}}' \right] & = - \nabla \hat{p}' + \Delta \hat{\vec{u}}', \\
\nabla \cdot \hat{\vec{u}}'& = 0.
\end{align}
To obtain these equations, we have used an estimate $l \Delta P / \mu \sim U^s$.
Reynolds and Strouhal numbers are defined as
\begin{equation}
Re = \frac{\rho U^s l}{\mu} \qquad \mbox{and} \qquad St = \frac{l}{t_s U^s},
\end{equation}
respectively. Here, $\rho$ is fluid density and $\mu$ is fluid dynamic viscosity.

\subsection{Fast macroscale flow decomposition} \label{app:mse-fastdecomp}

We employ so called
fast macroscale flow decomposition in the region, where there is only free
fluid. The decomposition for velocity and pressure is 
\begin{equation}
\hat{\vec{u}}' = \vec{U}' + \hat{\vec{u}}'^{+} \qquad \mbox{and} \qquad
\hat{p}' = P' + \hat{p}'^{+}, \label{appeq:fast-decomp}
\end{equation}
respectively. Here, $\hat{\vec{u}}'^{+}$ and $\hat{p}'^{+}$ are velocity and
pressure perturbation caused by textured or porous surface, while $\vec{U}'$
and $P'$ is flow field assumed to obey NS equations exposed to
homogeneous no-slip condition $\vec{U}' = 0$
at the artificial interface with the surface ($z = z_i$).
In the surface, we assume that
there is only slow flow denoted by $\hat{\vec{u}}'^{-}$ and $\hat{p}'^{-}$
satisfying same NS equations. Gathering resulting governing equations for
flow variables $\hat{\vec{u}}'^{\pm}$ and $\hat{p}'^{\pm}$ we obtain
\begin{align}
Re \left[ St\,\pdt \hat{\vec{u}}'^{+} + \left( \hat{\vec{u}}'^{+} \cdot \nabla \right)
\hat{\vec{u}}'^{+} + f\left( \hat{\vec{u}}'^{+}, \vec{U}' \right) \right] & = 
- \nabla \hat{p}'^{+} + \Delta \hat{\vec{u}}'^{+} & & z \geq z_i,
\label{appeq:ns-pert-1}\\
\nabla \cdot \hat{\vec{u}}'^{+}& = 0 & & z \geq z_i, \\
\ten{\Sigma^{u^-}}\cdot \vec{n} = \ten{\Sigma^{u^+}}\cdot \vec{n} +
\ten{\Sigma^{U}}\cdot \vec{n}, \ \ \ & \ \ \ 
\hat{\vec{u}}'^{-} = \hat{\vec{u}}'^{+} & & z = z_i, \\
Re \left[ St\,\pdt \hat{\vec{u}}'^{-} + \left( \hat{\vec{u}}'^{-} \cdot \nabla \right)
\hat{\vec{u}}'^{-} \right] & = - \nabla \hat{p}'^{-} + \Delta \hat{\vec{u}}'^{-}  & & z \leq z_i, \\
\nabla \cdot \hat{\vec{u}}'^{-}& = 0 & & z \leq z_i, \label{appeq:ns-pert-5}
\end{align}
where we have set continuity of velocity and stress at the artificial interface.
The forcing from macroscopic fast flow is defined as
\begin{equation}
f\left( \hat{\vec{u}}'^{+}, \vec{U}' \right) = 
\left( \vec{U}' \cdot \nabla \right) \hat{\vec{u}}'^{+} + 
\left( \hat{\vec{u}}'^{+} \cdot \nabla \right) \vec{U}',
\end{equation}
while the stress tensors are
\begin{equation}
\ten{\Sigma^{u^\pm}} = - \hat{p}'^{\pm}\,\ten{I} + \nabla \hat{\vec{u}}'^{\pm} +
\left( \nabla \hat{\vec{u}}'^{\pm} \right)^T \ \ \ \mbox{and} \ \ \
\ten{\Sigma^{U}} = - P'\,\ten{I} + \nabla \vec{U}' +
\left( \nabla \vec{U}' \right)^T.
\end{equation}
Taking sum of equations (\ref{appeq:ns-pert-1}--\ref{appeq:ns-pert-5}) with
governing equations for $\vec{U}'$ and $P'$, one recovers the single set of
equations for whole domain.

\subsection{Multi-scale expansion} \label{app:mse-expand}

%From now on, we turn to Einstein index notation for summation and 
%comma notation for differentiation.
To carry out the MSE, we introduce two different dimensionless
coordinates, so called macroscale and microscale, as
\begin{equation}
\vec{x}' = \frac{\vec{\hat{x}}}{H} \qquad \mbox{and} \qquad
\vec{\hat{x}}' = \frac{\vec{\hat{x}}}{l},
\end{equation}
respectively. Note that the second coordinate is identical as the one introduced
in the section~\ref{app:mse-nondim}. In these coordinates, there are two derivatives
appearing due to chain rule
\begin{equation}
\nabla = \nabla_1 + \ord \nabla_0,
\label{appeq:mse-chain}
\end{equation}
where $\nabla_1$ and $\nabla_0$ corresponds to derivatives
with respect to $\vec{\hat{x}}'$ and $\vec{x}'$, respectively.
Here, $\ord = l/H \ll 1$ is
scale separation parameter. In addition, the standard
amplitude expansion is employed for perturbation velocity and pressure fields
as
\begin{align}
\hat{\vec{u}}'^{\pm} & = \hat{\vec{u}}'^{\pm(0)} + \ord \hat{\vec{u}}'^{\pm(1)}
+ \ord^2 \hat{\vec{u}}'^{\pm(2)} + \ordest{\ord^3}, \label{appeq:mse-amp1} \\
\hat{p}'^{\pm} & = \hat{p}'^{\pm(0)} + \ord \hat{p}'^{\pm(1)}
+ \ord^2 \hat{p}'^{\pm(2)} + \ordest{\ord^3}. \label{appeq:mse-amp2}
\end{align}
We insert the chain rule (\ref{appeq:mse-chain}) and amplitude
expansions (\ref{appeq:mse-amp1}--\ref{appeq:mse-amp2}) in the governing
equations for perturbation velocity (\ref{appeq:ns-pert-1}--\ref{appeq:ns-pert-5}).
We assume that we are working with small Reynolds number $Re \leq \ordest{\ord^2}$
and small Strouhal number $St \leq \ordest{1}$ and group the terms appearing at
different orders.

\subsubsection{$\ordest{1}$ problem and solution Ansatz}

The problem at order $\ordest{1}$ reads
\begin{align}
- \nabla_1 \hat{p}'^{+(0)} + \Delta_1 \hat{\vec{u}}'^{+(0)} = 0 \ \ \  & \ \ \ 
\nabla_1 \cdot \hat{\vec{u}}'^{+(0)} = 0
 & & z \geq z_i,
\label{appeq:ord1-1}\\
\ten{\Sigma^{u^{-(0)}}}\cdot \vec{n} = \ten{\Sigma^{u^{+(0)}}}\cdot \vec{n} +
\ten{\Sigma^{U}}\cdot \vec{n}, \ \ \ & \ \ \ 
\hat{\vec{u}}'^{-(0)} = \hat{\vec{u}}'^{+(0)} & & z = z_i, \label{appeq:ord1-2} \\
- \nabla_1 \hat{p}'^{-(0)} + \Delta_1 \hat{\vec{u}}'^{-(0)} = 0 \ \ \ &  \ \ \ 
\nabla_1 \cdot \hat{\vec{u}}'^{-(0)} = 0 & & z \leq z_i. \label{appeq:ord1-3}
\end{align}
One can observe, that this problem is forced by stress at the interface
from the fast macroscopic flow $\vec{U}'$. Therefore we anticipate that the
solution for velocity perturbations will take form
\begin{equation}
\hat{\vec{u}}'^{\pm(0)} = \ten{L'}_e^\pm \cdot \vec{\tau}'_e, \label{appeq:ord1-a1}
\end{equation}
where $\ten{L'}_e$ is unknown $3 \times 3$ tensor field,
and $\vec{\tau}'_e = \ord \left(\nabla_0 \vec{U}' + 
\left( \nabla_0 \vec{U}' \right)^T\right) \cdot \vec{n}$ is viscous stress vector
at the surface. The pressure, on the other hand, should take form
\begin{equation}
\hat{p}'^{+(0)} = \vec{b'}^+ \cdot \vec{\tau}'_e \ \ \mbox{and} \ \ 
\hat{p}'^{-(0)} = \vec{b'}^- \cdot \vec{\tau}'_e + p'^{-(0)}, \label{appeq:ord1-a2}
\end{equation}
where $\vec{b'}^\pm$ is unknown vector field
and $p'^{-(0)}$ is zeroth order pressure field below the interface containing only
macroscale variations. This term is needed to match the free flow pressure $P'$.
%The prefactors in Ansatzes (\ref{appeq:ord1-a1}--\ref{appeq:ord1-a2}) can be determined
%by inserting those into equations (\ref{appeq:ord1-1}--\ref{appeq:ord1-3}) and solving
%the resulting problems. The velocity condition is then obtained using a surface
%average of velocity Ansatz (\ref{appeq:ord1-a1}), giving slip boundary condition as the
%leading order approximation. The pressure jump condition is obtained
%as difference between volume averages of pressure Ansatz (\ref{appeq:ord1-a2})
%in free fluid side and surface side. This gives the
%pressure jump proportional to shear stress as leading order approximation.

\subsubsection{$\ordest{\ord}$ problem and solution Ansatz}

The problem at order $\ordest{\ord}$ is obtained by collecting all the terms
at appropriate order and inserting solution of fields from the order $\ordest{1}$
problem. Resulting equations are
\begin{align}
- \nabla_1 \hat{p}'^{+(1)} + \Delta_1 \hat{\vec{u}}'^{+(1)} & 
= g\left(\ten{L'}_e^+, 
\vec{b'}^+\right) : \nabla_0 \vec{\tau}'_e  & & z \geq z_i, \label{appeq:ord2-1} \\
\nabla_1 \cdot \hat{\vec{u}}'^{+(1)} & = - \ten{L'}_e^+ : \nabla_0 \vec{\tau}'_e
 & & z \geq z_i, \\
\ten{\Sigma^{u^{-(1)}}}\cdot \vec{n} = \ten{\Sigma^{u^{+(1)}}}\cdot \vec{n},
 \ \ \ & \ \ \ 
\hat{\vec{u}}'^{-(1)} = \hat{\vec{u}}'^{+(1)} & & z = z_i, \\
- \nabla_1 \hat{p}'^{-(1)} + \Delta_1 \hat{\vec{u}}'^{-(1)} & 
= g\left(\ten{L'}_e^-, 
\vec{b'}^-\right) : \nabla_0 \vec{\tau}'_e + \nabla_0 p'^{-(0)}  & & z \leq z_i, \\
\nabla_1 \cdot \hat{\vec{u}}'^{-(1)} & = - \ten{L'}_e^- : \nabla_0 \vec{\tau}'_e
 & & z \leq z_i, \label{appeq:ord2-3}
\end{align}
where we the double contraction
between $3 \times 3 \times 3$ tensor $\ten{A}$ and $3 \times 3$ tensor
$\ten{B}$ is defined as $\ten{A} : \ten{B} = A_{ijk} B_{jk}$.
Here one observes that the problem is forced using volume forcing and
mass sources which are proportional to macroscopic gradient of
pressure within the surface as well as macroscopic gradient of free flow
shear stress. The volume forcing factor standing in front of shear stress
variations is defined as $ g\left(\ten{L'}_e^\pm, 
\vec{b'}^\pm\right) = \vec{b'}^\pm \ten{\delta} + 2 \nabla_1 \ten{L'}_e^\pm$.
Therefore we assume that the
solution for next order velocity perturbations is
\begin{equation}
\hat{\vec{u}}'^{\pm(1)} = - \ten{K'}_e^\pm \cdot \nabla_0 p'^{-(0)}
- \ten{M'}_e^\pm : \nabla_0 \vec{\tau}'_e, \label{appeq:ord2-a1}
\end{equation}
where $\ten{K'}_e^\pm$ is unknown $3 \times 3$ tensor field,
and $\ten{M'}_e^\pm$ is unknown $3 \times 3 \times 3$ tensor
field. The pressure, on the other hand, should take the following form
\begin{equation}
\hat{p}'^{\pm(1)} = - \vec{a'}^\pm \cdot \nabla_0 p'^{-(0)}
- \ten{C'}^\pm : \nabla_0 \vec{\tau}'_e,  \label{appeq:ord2-a2}
\end{equation}
where $\vec{a'}^\pm$ is an unknown vector field and $\ten{C'}^\pm$ is unknown
$3 \times 3$ tensor field.
%The prefactors in Ansatzes (\ref{appeq:ord2-a1}--\ref{appeq:ord2-a2}) can be determined
%by inserting those into equations (\ref{appeq:ord2-1}--\ref{appeq:ord2-3}) and solving
%the resulting problems. The correction for velocity condition is
%then obtained using a surface
%average of velocity Ansatz (\ref{appeq:ord2-a1}), giving Darcy-type velocity and
%velocity proportional to variation of shear stress as the
%next order correctors. The next order pressure jump condition is obtained
%as difference between volume averages of pressure Ansatz (\ref{appeq:ord2-a2})
%in free fluid side and surface side. This gives the next order correction of
%pressure jump proportional to pressure gradient and variation of shear stress.

\subsection{Boundary conditions up to order $\ordest{\ord}$}

In this section, we present resulting boundary conditions.
% illustrate how Ansatzes (\ref{appeq:ord1-a1}--\ref{appeq:ord1-a2},
%\ref{appeq:ord2-a1}--\ref{appeq:ord2-a2}) can be used to deduce the form of boundary
%condition between a textured or porous surface and free fluid
 The result and derivation
holds both for textured and porous surface.
%, with exception that for the textured surface
%there is no 

\subsubsection{Order $\ordest{1}$ boundary conditions} \label{appsec:ord1-bc}

To determine the boundary condition with an error of $\ordest{\ord}$,
we use Ansatzes for solution of $\ordest{1}$-problem
(\ref{appeq:ord1-a1}--\ref{appeq:ord1-a2}) and insert them back into the
amplitude expansion (\ref{appeq:mse-amp1}--\ref{appeq:mse-amp2}).

For velocity condition, we take the surface average at the interface, neglect
all higher order terms, relate the no-slip solution with the corrected
flow field and go back to the dimensional quantities to obtain
\begin{equation}
\vec{u} = \left( l\, \ten{\mathcal{L}} \right) \cdot \frac{\vec{\tau}_e}{\mu}
+ \ordest{\ord\,U^s} \qquad \mbox{on} \ \ z = z_i,
\end{equation}
where the dimensionless $3 \times 3$ tensor $\ten{\mathcal{L}} = 
\langle \ten{L}'^+_e \rangle_i$ is a
surface average of the
microscale tensor field $\ten{L}'^+_e$.

For pressure jump condition, we follow the same approach as for velocity,
but instead of single surface average we take instead two volume averages
(one in the free fluid region, and second in the interior region) and take
a difference for estimating the pressure jump condition. For the estimation
of the pressure jump, we have to remember also that the pressure in the
free fluid contain both the perturbation and the no-slip
solution (\ref{appeq:fast-decomp}).
Taking all this into account, we obtain
\begin{equation}
p^- - p = \vec{\bar{b}} \cdot \vec{\tau}_e
+ \ordest{\ord\,\Delta P} \qquad \mbox{on} \ \ z = z_i,
\end{equation}
where vector with $3$ components $\vec{\bar{b}}
= \langle \vec{b}'^- \rangle^-
- \langle \vec{b}'^+ \rangle^+$ is the difference in volume averaged fields
$\vec{b}^\pm$. 

\subsubsection{Order $\ordest{\ord}$ boundary conditions} \label{appsec:ord2-bc}

For boundary condition with an error of $\ordest{\ord^2}$,
we repeat the same procedure as in section~\ref{appsec:ord1-bc}
by taking additionally into account the
Ansatzes for solution of $\ordest{\ord}$-problem
(\ref{appeq:ord2-a1}--\ref{appeq:ord2-a2}).

Velocity boundary condition then becomes
\begin{equation}
\vec{u} = \left( l\, \ten{\mathcal{L}} \right) \cdot \frac{\vec{\tau}_e}{\mu}
- \left(l^2\,\ten{\mathcal{K}} \right) \cdot \frac{\nabla p^-}{\mu}
+ \left(l^2\,\ten{\mathcal{M}} \right) : \frac{\nabla\vec{\tau}_e}{\mu}
+ \ordest{\ord^2\,U^s} \qquad \mbox{on} \ \ z = z_i, \label{appeq:bc-mse-vel}
\end{equation}
where the dimensionless $3 \times 3$ tensor $\ten{\mathcal{K}} = 
\langle \ten{K}'^+_e \rangle_i$ is a
surface average of the
microscale tensor field $\ten{K}'^+_e$ and the dimensionless 
$3 \times 3 \times 3$ tensor $\ten{\mathcal{M}} = \langle \ten{M}'^+_e \rangle_i$
is the surface average of the microscale tensor field $\ten{M}'^+_e$.

%The tensor
%$\ten{\mathcal{L}}$ provides zero contribution for interface normal velocity $u_n$
%as before, while tensor $\ten{\mathcal{K}}$ gives contribution to interface normal
%velocity $u_n$ for porous materials only.
%Here, the term with tensor $\ten{\mathcal{M}}$ can be related to transpiration
%length, after applying the relationship between tangential derivatives
%of shear stress and leading order slip velocity boundary condition. 
%Note that the approach of using first derivatives of slip velocity
%is possible only for the reduced set of terms, because it is not possible
%to differentiate boundary condition in interface normal direction.

The pressure boundary condition with next order corrections becomes
\begin{equation}
p^- - p = \vec{\bar{b}} \cdot \vec{\tau}_e
- \left( l\, \vec{\bar{a}} \right) \cdot \nabla p^-
+ \left( l\, \ten{\mathcal{C}} \right) : \nabla \vec{\tau}_e
+ \ordest{\ord^2\,\Delta P} \qquad \mbox{on} \ \ z = z_i, \label{appeq:bc-mse-p}
\end{equation}
where dimensionless vector with $3$ components $\vec{\bar{a}}
= \langle \vec{a}'^- \rangle^-
- \langle \vec{a}'^+ \rangle^+$ is the difference in volume averaged fields
$\vec{a}^\pm$ and dimensionless $3 \times 3$ tensor $\ten{\mathcal{C}}
= \langle \ten{C}'^- \rangle^-
- \langle \ten{C}'^+ \rangle^+$ is the difference in volume averaged fields
$\ten{C}'^\pm$. The boundary
conditions (\ref{appeq:bc-mse-vel}--\ref{appeq:bc-mse-p}) are presented in
the main paper (\ref{eq:vel-bc-from-mse}--\ref{eq:p-bc-from-mse}) and discussed
in the context of the TR model.

%\section{Dirac delta surface forcing} \label{app:dirac-delta}
\section{Equivalence to a two-domain description} \label{app:dirac-delta}

In this appendix, we elaborate on how the Dirac delta function
is used for surface forcing in
equations (\ref{eq:itf-cell-1}--\ref{eq:itf-cell-2}). In essence, this
notation is equivalent to having a two-domain description and enforcing
continuity of velocities and jump in stress, as appearing in multi-scale expansion
(\ref{appeq:ord1-2}) and also as reported
in work by \cite{lacis2016}.

Let us consider the equations (\ref{eq:itf-cell-1}--\ref{eq:itf-cell-2}) in
three different regions; (i) above the interface, (ii) below the interface
and (iii) in a close vicinity of the interface. Introducing plus notation
for variables above the interface and minus notation for variables below
the interface, we rewrite (\ref{eq:itf-cell-1}--\ref{eq:itf-cell-2}) as
\begin{align}
- \nabla \hat{p}^+ + \mu \Delta \hat{\vec{u}}^+ & = 0, &
  \nabla \cdot \hat{\vec{u}}^+ = 0, & & \hat{z} > \hat{z}_i, \\
- \nabla \hat{p}^- + \mu \Delta \hat{\vec{u}}^- & = 0, &
  \nabla \cdot \hat{\vec{u}}^- = 0, & & \hat{z} < \hat{z}_i,
\end{align}
where we have used the property of the Dirac delta that it is zero everywhere
except at $\hat{z} = \hat{z}_i$. Now, however, there is additional surface that requires
new boundary conditions. From the continuity of solution at the single domain,
we state that the first condition at $\hat{z} = \hat{z}_i$ must be continuity of velocities,
i.e., $\hat{\vec{u}}^+ = \hat{\vec{u}}^-$. This is, however, not sufficient and a
stress condition is also required.
Consider equations (\ref{eq:itf-cell-1}--\ref{eq:itf-cell-2}) in the near
vicinity of the interface $\hat{z} = \hat{z}_i$ before introduction of plus and minus
notation.
We rewrite the momentum equation (\ref{eq:itf-cell-1}) by making use of the
Newtonian fluid stress tensor $\ten{\Sigma}$ as
\begin{equation}
\nabla \cdot \left[ - \hat{p} \ten{I} + \mu \left\{ \nabla \vec{\hat{u}} 
+ \left( \nabla \vec{\hat{u}} \right)^T \right\} \right] = \nabla \cdot \ten{\Sigma}
= - \delta \left(\hat{z} -
\hat{z}_i \right) \vec{\tau},
\end{equation}
where $\ten{I}$ is identity tensor.
We integrate the equation in the interface
normal direction from $\hat{z}_i - \delta \hat{z}$ to $\hat{z}_i + \delta \hat{z}$ and get
\begin{equation}
\left. \ten{\Sigma} \right|_{\hat{z}_i + \delta \hat{z}} \cdot \vec{e}_z
 - \left. \ten{\Sigma} \right|_{\hat{z}_i - \delta \hat{z}} \cdot \vec{e}_z
+ \int\limits_{\hat{z}_i - \delta \hat{z}}^{\hat{z}_i + \delta \hat{z}} \left[ \pd_x \ten{\Sigma} \cdot \vec{e}_x
 + \pd_y \ten{\Sigma} \cdot \vec{e}_y
\right]\,\mathit{d\hat{z}} = - \vec{\tau}, \label{appeq:moment-integr}
\end{equation}
where we have explicitly integrated out the divergence part in $z$-direction as well as Dirac delta
function, which gives one as long as the integration interval is encapsulating the interface coordinate
from both sides. Assuming that the fluid stress tensor varies smoothly in $x$ and $y$ directions, we
have
\begin{equation}
\int\limits_{\hat{z}_i - \delta \hat{z}}^{\hat{z}_i + \delta \hat{z}} \left[ \pd_x \ten{\Sigma} \cdot \vec{e}_x
 + \pd_y \ten{\Sigma} \cdot \vec{e}_y
\right]\,\mathit{d\hat{z}} = 0
\end{equation}
as the integration interval shrinks to zero $\delta \hat{z} \rightarrow 0$.
Consequently the moment equation
integral (\ref{appeq:moment-integr}) can be rewritten as
\begin{equation}
\left. \ten{\Sigma} \right|_{\hat{z}_i^-} \cdot \vec{e}_z = \left. \ten{\Sigma} \right|_{\hat{z}_i^+} \cdot \vec{e}_z + \vec{\tau},
\label{appeq:stress-jump-pre}
\end{equation}
where fluid stress tensors are evaluated at the interface from the negative
%side $\left. \ten{\Sigma} \right|_{z_i^-} = - \hat{p}^- \ten{I} + \mu \left\{ \nabla \vec{\hat{u}}^-
%+ \left( \nabla \vec{\hat{u}}^- \right)^T \right\}$ and from the positive
%side $\left. \ten{\Sigma} \right|_{z_i^+} = - \hat{p}^+ \ten{I} + \mu \left\{ \nabla \vec{\hat{u}}^+
%+ \left( \nabla \vec{\hat{u}}^+ \right)^T \right\}$.
side $\left. \ten{\Sigma} \right|_{\hat{z}_i^-}$ (function
of $\hat{p}^-$ and $\vec{\hat{u}}^-$) and from the positive
side $\left. \ten{\Sigma} \right|_{\hat{z}_i^+}$ (function 
of $\hat{p}^+$ and $\vec{\hat{u}}^+$).
By taking into account the
orientation of unit normal of the interface $\vec{e}_z = \vec{n}$, we have
\begin{equation}
\left. \ten{\Sigma} \right|_{\hat{z}_i^-} \cdot \vec{n} = \left. \ten{\Sigma} \right|_{\hat{z}_i^+} \cdot \vec{n} + \vec{\tau},
\label{appeq:stress-jump}
\end{equation}
which is the final boundary condition needed for the two domain formulation of the
interface cell problem.

\section{Description of numerical methods} \label{app:sim-detail}

In this appendix, we describe more details of the numerical methods we have
used through this work.

\subsection{Laminar flow} \label{app:lam-detail}

To discretise the domain for the rough configuration (figure~\ref{fig:cavity-rough-geom}),
we use node spacing $\Delta s_t = 0.05\,l$ at the top wall
and $\Delta s_b = 0.005\,l$ at the surface texture.
%%%%%We have checked convergence of DNS by using $\Delta s_t = 0.25\,l$ and $\Delta s_b = 0.025\,l$ (fivefold increase of mesh spacing) and observed that velocities near the surface texture changed by less than $1\%$.
The very fine mesh for the rough configuration was chosen to make sure that the variations
in the final averaged data are not due to resolution issues.
The node spacing for effective textured simulations is 
$\Delta s_t = \Delta s_b = 0.05\,l$.

To discretise the domain for the porous configurations (figure~\ref{fig:geom}),
we use node spacing $\Delta s_t = 0.125\,l$ at the top wall
and $\Delta s_b = 0.05\,l$ at the porous structures.
%We have checked convergence by using $\Delta s_t = 0.0555\,l$ and $\Delta s_b = 0.033\,l$ (around $40\%$ reduction in mesh spacing) and observed that change of velocities near the porous region were always smaller  than $0.2\%$.
For effective simulations of porous configurations, we use node
spacing $\Delta s_t = \Delta s_b = 0.083\,l$ at all walls. 
%We have checked convergence by using $\Delta s_t = \Delta s_b = 0.125\,l$ ($50\%$ reduction in mesh size) and observed that change in velocities near the porous material were always below $0.4\%$.

We %use standard techniques to solve the set of 
solve the incompressible Stokes equations
with finite element solver FreeFEM++ \citep{freefem}. We choose a monolithic approach,
i.e., the momentum and continuity equations are treated at the same time, which leads
to natural treatment of boundary conditions, which mix velocities and pressures.

\subsection{Turbulent flow} \label{app:turb-detail}

For turbulent simulations, we
use a periodic uniform grid in the spanwise ($x$) and streamwise ($y$) directions, and a stretched grid in wall normal ($z$) direction. The solver is based on a staggered grid with a third order Runge-Kutta time scheme combined with a splitting technique. A semi-explicit sub-iteration scheme is used to implement the model velocity boundary conditions.
The results are made dimensionless using the viscous length and the friction velocity defined as $u_{\tau} / \nu$ and $u_{\tau}$, respectively.  These quantities are computed at the cuboids crest plane to be compared to the TR model. The mesh spacings for the geometry resolving simulation in viscous units are $\Delta x^+ = \Delta y^+ = 1.794$, $\Delta z_w^+ = 0.358$, and $\Delta z_c^+ = 3.096$; subscripts 'w' and 'c' denote near-wall and channel center line respectively. Within the textured layer, we use a constant near-wall mesh spacing $\Delta z_w^+$. We also impose 5 layers of uniform $\Delta z_w^+$ at both channel walls.  For the effective (and smooth wall) simulations, the mesh spacing is $\Delta x^+=7.8$, $\Delta y^+=5.1$, $\Delta z_w^+=0.3$, and $\Delta z_c^+=4.8$.

The Reynolds number for the geometry-resolved case is defined as $Re_{rough} =  U (\delta+k/2) / \nu = 2856$ based on the bulk velocity $U=1$, the half-channel height and the kinematic viscosity $\nu$. As the height of the domain is truncated for the TR model, we consider the Reynolds number computed on the reduced domain, cutting off the roughness part. It becomes $Re = U \delta / \nu = 2839.2$ with $U_c = 1.014$ the bulk velocity computed on this truncated domain. This Reynolds number is kept constant for all DNSs, which leads to $Re_{\tau} =u_\tau\delta/\nu \approx 180$ for smooth channel, where $u_\tau$ is the friction velocity.

\bibliographystyle{jfm}
\bibliography{porous}

\end{document}